\documentclass[review,12pt]{elsarticle}

\usepackage{amssymb}
\usepackage{amsthm}
\usepackage{framed} 
\usepackage{amsmath,color}
\usepackage{mathrsfs}
\usepackage{graphicx}
\usepackage{epstopdf}
\usepackage{float}
\usepackage{caption}
\usepackage{subcaption}
\usepackage{bm}
\usepackage{bbm}
\usepackage{mathrsfs}
\usepackage{hyperref}
\usepackage{cleveref}
\usepackage{soul}
\usepackage{svg} 
\usepackage{accents}
\usepackage{color,soul} 
\usepackage{color} 
\usepackage{bm}
\usepackage{multirow} 
\usepackage[margin=2cm]{geometry}
\usepackage{adjustbox}
\usepackage{comment}
\usepackage{blindtext}
\biboptions{sort&compress}
\soulregister\citep7 
\soulregister\citet7 
\soulregister\citealp7 
\newsavebox{\measurebox} 
\usepackage{titlesec} 
\usepackage[T1]{fontenc} 
\usepackage{lmodern} 
\input{glyphtounicode}  

\def\onedot{$\mathsurround0pt\ldotp$}
\def\cdddot#1{
  \mathbin{\vcenter{\baselineskip.67ex
    \hbox{\onedot}\hbox{\onedot}\hbox{\onedot}%
  }}%
}

\journal{Engineering Fracture Mechanics}

\makeatletter
\def\@author#1{\g@addto@macro\elsauthors{\normalsize%
    \def\baselinestretch{1}%
    \upshape\authorsep#1\unskip\textsuperscript{%
      \ifx\@fnmark\@empty\else\unskip\sep\@fnmark\let\sep=,\fi
      \ifx\@corref\@empty\else\unskip\sep\@corref\let\sep=,\fi
      }%
    \def\authorsep{\unskip,\space}%
    \global\let\@fnmark\@empty
    \global\let\@corref\@empty  
    \global\let\sep\@empty}%
    \@eadauthor={#1}
}
\makeatother

\titleformat{\paragraph}
{\normalfont\normalsize\itshape}{\theparagraph}{1em}{}
\titlespacing*{\paragraph}
{0pt}{3.25ex plus 1ex minus .2ex}{1.5ex plus .2ex}

\graphicspath{{Figures/}}

\begin{document}

\begin{frontmatter}



\title{Damage Mechanics Challenge: Predictions based on the phase field fracture model}


\author{Yousef Navidtehrani\fnref{Uniovi}}
\author{Ravindra Duddu \fnref{Vanderbilt}}
\author{Emilio Mart\'{\i}nez-Pa\~neda\corref{cor1}\fnref{IC,OXFORD}}
\ead{emilio.martinez-paneda@eng.ox.ac.uk}

\address[Uniovi]{Department of Construction and Manufacturing Engineering, University of Oviedo, Gij\'{o}n 33203, Spain}

\address[Vanderbilt]{Department of Civil and Environmental Engineering and Department of Mechanical Engineering, Vanderbilt University, PMB 351831, Nashville, TN 37235-1831, USA}

\address[IC]{Department of Civil and Environmental Engineering, Imperial College London, London SW7 2AZ, UK}

\address[OXFORD]{Department of Engineering Science, University of Oxford, Oxford OX1 3PJ, UK}

\cortext[cor1]{Corresponding author.}

\begin{abstract}
In this work, we describe our contribution to the Purdue-SANDIA-LLNL \emph{Damage Mechanics Challenge}. The phase field fracture model is adopted to blindly estimate the failure characteristics of the challenge test, an unconventional three-point bending experiment on an additively manufactured rock resembling a type of gypsum. The model is formulated in a variationally consistent fashion, incorporating a volumetric-deviatoric strain energy decomposition, and the numerical implementation adopts a monolithic unconditionally stable solution scheme. Our focus is on providing an efficient and simple yet rigorous approach capable of delivering accurate predictions based solely on physical parameters. Model inputs are Young's modulus $E$, Poisson's ratio $\nu$, toughness $G_c$ and strength $\sigma_c$ (as determined by the choice of phase field length scale $\ell$). We show that a single mode I three-point bending test is sufficient to calibrate the model, and that the calibrated model can then reliably predict the force versus displacement responses, crack paths and surface crack morphologies of more intricate three-point bending experiments that are inherently mixed-mode. Importantly, our peak load, crack trajectory and crack surface morphology predictions for the challenge test, submitted before the experimental data was released, show a remarkable agreement with experiments. The characteristics of the challenge, and how changes in these can impact the predictive abilities of phase field fracture models, are also discussed.\\
\end{abstract}

\begin{keyword}

Phase field fracture \sep Damage mechanics challenge \sep Brittle fracture \sep Rock fracture \sep Finite element analysis



\end{keyword}

\end{frontmatter}



\section{Introduction}
\label{Sec:Introduction}

This work presents our contribution to the \emph{Damage Mechanics Challenge} organised by Purdue University, Sandia National Laboratories and the Lawrence Livermore National Laboratory. The aim of the \emph{Damage Mechanics Challenge} is to assess and showcase the ability of computational methods to predict (as opposed to fit) the failure of rock-like materials. As detailed below, a certain degree of information was provided on the deformation and fracture characteristics of the material under consideration, a 3D-printed rock, and then participants were asked to predict - using the computational approach of their choosing - the failure behaviour (force vs displacement response, crack trajectory and morphology) in a new test configuration.\\

We chose to employ the phase field fracture model \cite{Francfort1998,Bourdin2008} to predict the fracture behaviour of the additively manufactured rock samples, due to its robustness and rigorous physical basis. The phase field fracture model has been enjoying an ever-growing popularity in recent years. Grounded on Griffith's energy balance \cite{Griffith1920}, the phase field model enables predicting complex cracking phenomena based on the thermodynamics of fracture, including arbitrary crack branching, coalescence and arbitrary crack trajectories. The approach is also known to be mesh objective and computationally robust \cite{Miehe2010a,Kristensen2020}. Hence, not surprisingly, phase field-based fracture models have been developed to simulate material failure across a wide range of engineering applications, including dynamic fracture \cite{Borden2012,McAuliffe2016,Mandal2020}, hydrogen embrittlement \cite{CMAME2018,Cui2022}, fatigue damage \cite{Carrara2020,Golahmar2023}, fibre-reinforced composites \cite{Quintanas-Corominas2019,CST2021,Gao2023}, functionally graded materials \cite{CPB2019,Kumar2021}, smart materials \cite{CMAME2021,Simoes2022,Quinteros2022}, and Li-Ion battery degradation \cite{Klinsmann2016,Boyce2022}. Notably, phase field fracture methods have been recently applied to study crack propagation in rock-like materials \cite{Liu2022,Hug2022,Xu2022,Bryant2018,Santillan2017} and glaciers \cite{Sun2021,Clayton2022}.\\

In the following, we proceed to describe how we have successfully predicted the required experimental outcome of the \emph{Damage Mechanics Challenge} using the phase field fracture model and the provided experimental calibration data. The aim was to demonstrate robust predictive capabilities with minimal model complexity. As such, a conventional phase field fracture model (so-called \texttt{AT2} model \cite{Bourdin2008}) is employed, such that predictions depend only on four material properties: Young's modulus $E$, Poisson's ratio $\nu$, fracture energy $G_c$, and strength $\sigma_c$, with the last one being indirectly defined through the choice of phase field length scale $\ell$. It is worth emphasising that this contribution deals with a piece of work that was conducted as part of the standard \emph{Damage Mechanics Challenge}; i.e., the results presented are \emph{blind} predictions, which were submitted to the challenge organisers before the experimental data of the benchmark test was released. 

\section{Approach: Phase field fracture modelling}
\label{Sec:2approach}

In the following, we proceed to describe the numerical approach employed and the characteristics of the boundary value problem under consideration. This study was carried out by three researchers based at the University of Oviedo, University of Oxford, and Vanderbilt University with previous collaborative experience in related endeavours. 

\subsection{A phase field description of fracture}

\subsubsection{Background}

The phase field fracture model builds upon Griffith's foundational thermodynamic framework \cite{Griffith1920}. In concordance with the principles of the first law of thermodynamics, the initiation or propagation of a crack is contingent upon the proviso that the total energy of the system either diminishes or remains constant. Thus, the condition for fracture is critically determined through equilibrium considerations, whereby the overall energy remains unaltered. Consider an elastic solid including a crack, the perturbation in the total energy $\mathcal{E}$ attributable to infinitesimal growth in the crack area, denoted as d$A$, can be articulated as:
\begin{equation}\label{eq:Egriffith0}
\frac{\text{d}\mathcal{E}}{\text{d}A} = \frac{\text{d} \Pi }{\text{d} A} + \frac{\text{d}W_c}{\text{d}A} =  \frac{\text{d} \Psi \left( \bm{\varepsilon} \left( \mathbf{u} \right) \right) }{\text{d} A} + \frac{\text{d}W_e}{\text{d}A} + \frac{\text{d}W_c}{\text{d}A} = 0
\end{equation}

\noindent where $W_c$ denotes the energy expenditure required to generate two new surfaces, and $\Pi$ is the total potential energy supplied by the internal strain energy $\Psi$ and the external forces $W_e$. The last term in Eq. (\ref{eq:Egriffith0}) is the so-called fracture energy or critical energy release rate, $G_c=\text{d}W_c/\text{d}A$; a constant, material-specific parameter characterising its resilience against fracture. The internal strain energy $\Psi$ is a function of the strain field $\bm{\varepsilon}$, which is itself a function of the displacement field; for small strains, $\bm{\varepsilon}=\left( \nabla \mathbf{u}^T + \nabla \mathbf{u} \right)/2$. Thus, in the case of prescribed/fixed displacements, although no external work is done on the body ($W_e=0$), a crack would grow if the energy stored in the solid equates to the energy required to create two new surfaces. As such, Griffith's hypothesis describes a localised principle of minimality governing the cumulative stored and fracture energies. Within an arbitrary domain $\Omega \subset {\rm IR}^n$ $(n \in[1,2,3])$ encompassing an internal discontinuity boundary $\Gamma$, this principle of minimality can be expressed through a variational representation as:
\begin{equation}\label{eq:Egriffith}
\mathcal{E} \left( \mathbf{u} \right) = \int_\Omega \psi \left( \bm{\varepsilon} \left( \mathbf{u} \right) \right) \text{d}V + \int_\Gamma G_c \, \text{d}S - \int_\Omega  \mathbf{b} \cdot \mathbf{u} \, \text{d}V - \int_{\partial \Omega}  \mathbf{T} \cdot \mathbf{u} \, \text{d}S \,
\end{equation}

\noindent where the external work is characterised by the body force $\mathbf{b}$ and the external traction vector $\mathbf{T}$, and their dot product with the displacement vector $\mathbf{u}$. Thus, the trajectory of crack growth can be predicted devoid of arbitrary criteria, grounded in the principles of global minimality and the conversion of stored energy into fracture energy. Nonetheless, minimising the Griffith energy functional (\ref{eq:Egriffith}) is hindered by the intricacies associated with the tracking of the advancing fracture surface $\Gamma$. This computational challenge can be addressed by making use of a scalar phase field variable $\phi$, which can be interpreted as a damage field variable, transitioning from 0 in undamaged regions to 1 within the confines of the crack. In alignment with the rationale of continuum damage mechanics, a degradation function $g(\phi)=(1-\phi)^2$ is also used, so as to modulate the material stiffness in accordance with the evolving damage. Consequently, the regularised energy functional takes the form:
\begin{equation}\label{eq:EgriffithR}
\mathcal{E}_{\ell} \left( \mathbf{u}, \phi \right) = \int_\Omega \left( 1 - \phi \right)^2 \psi_0 \left( \bm{\varepsilon} \left( \mathbf{u} \right) \right) \text{d}V + \int_\Omega G_c \gamma_\ell \left( \phi \right) \text{d}V  - \int_\Omega  \mathbf{b} \cdot \mathbf{u} \, \text{d}V - \int_{\partial \Omega}  \mathbf{T} \cdot \mathbf{u} \, \text{d}S \,
\end{equation}

\noindent where $\gamma_\ell$ is the so-called crack density function, which for the conventional \texttt{AT2} model reads \cite{Bourdin2008}:
\begin{equation}\label{eq:CrackDensityFunction}
\gamma_\ell \left( \phi \right) = \frac{\phi^2}{2 \ell} + \frac{\ell}{2} |\nabla \phi|^2.
\end{equation}

\noindent The crack density function includes the gradient of the phase field order parameter and accordingly a length scale $\ell$, which enables mesh objectivity. This phase field length scale $\ell$ is directly related to the material strength, as can be illustrated by plotting the solution to the homogeneous, 1D coupled deformation-phase field fracture problem, which gives a maximum stress of,
\begin{equation}
\sigma_c = \sqrt{\frac{27 E G_c}{256 \ell}}
\label{eq:Hom-Solution}
\end{equation}

Accordingly, for plane stress conditions, $\sigma_c \propto \sqrt{G_c E /\ell} = K_{Ic}/\sqrt{\ell}$, and the choice of $\ell$ will define the material strength for a given Young's modulus $E$ and critical fracture energy $G_c$ (or fracture toughness $K_{Ic}$). The ability of the phase field fracture model to go beyond Griffith's fracture and incorporate the concept of material strength is essential to predict crack nucleation \cite{Tanne2018}, and as a result, the phase field fracture model can capture the transition from toughness-driven failures to strength-driven failures \cite{PTRSA2021}, naturally encompassing the transition flaw size concept. Thus, the structure of the phase field model ensures its alignment with conventional fracture mechanics theory and this has been shown computationally (see, e.g. Refs. \cite{Klinsmann2015,Tanne2018,PTRSA2021}) and theoretically - e.g., $\Gamma$-convergence studies have shown that the regularised functional (\ref{eq:EgriffithR}) converges to the Griffith functional (\ref{eq:Egriffith}) in both discrete and continuous systems \cite{Bellettini1994,Chambolle2004}.\\

One relevant aspect to consider is that the conventional phase field fracture model, akin to Griffith's work, assumes a symmetric fracture behaviour in tension and compression. To break this symmetry and hinder cracking in compressive regions, a number of authors have proposed modifications to the model that aim at decomposing the fracture driving force, the strain energy density. Accordingly, the (undamaged) strain energy density, which is typically defined as follows for elastic solids,
\begin{equation}\label{eq:consti}
\psi_0 (\bm{\varepsilon} \left( \mathbf{u}  \right)) = \frac{1}{2} \bm{\varepsilon} \left( \mathbf{u}  \right) : \bm{C}_0 : \bm{\varepsilon} \left( \mathbf{u}  \right),
\end{equation}
with $\bm{C}_0$ denoting the undamaged stiffness tensor, can be decomposed into a tensile part, $\psi_0^+$, and a compressive part, $\psi_0^-$, such that
\begin{equation}\label{eqn:str-decomposition}
\psi_0 (\bm{\varepsilon} (\mathbf{u})) = \psi_0^+ (\bm{\varepsilon} (\mathbf{u})) + \psi_0^- (\bm{\varepsilon} (\mathbf{u})),
\end{equation}

Herein, we adopt the so-called volumetric-deviatoric split \cite{Amor2009}, rendering
\begin{equation}\label{Eq:Split1}
\begin{aligned}
& \psi_0^{+} (\varepsilon (\mathbf{u}))=\frac{1}{2} K\langle\operatorname{tr}(\boldsymbol{\varepsilon} (\mathbf{u}))\rangle_{+}^2+\mu\left(\varepsilon^{\prime} (\mathbf{u}): \boldsymbol{\varepsilon}^{\prime} (\mathbf{u})\right) \ \\
& \psi_0^{-} (\varepsilon (\mathbf{u}))=\frac{1}{2} K\langle\operatorname{tr}(\varepsilon (\mathbf{u}))\rangle_{-}^2,
\end{aligned}
\end{equation}

\noindent where $K$ denotes the bulk modulus, $\mu$ represents the shear modulus, and $\langle \rangle$ are used to denote the Macaulay brackets, defined as $\langle a \rangle_{\pm}=(a\pm |a|)/2$. Furthermore, $\bm{\varepsilon}' (\mathbf{u})$ is the deviatoric part of the strain tensor, defined as $\bm{\varepsilon}' (\mathbf{u})=\bm{\varepsilon} (\mathbf{u})-\mathrm{tr}(\bm{\varepsilon} (\mathbf{u})) \bm{1} /3$, where $\bm{1}$ denotes the second-order unit tensor.

\subsubsection{Balance equations}

We now proceed to formulate the relevant partial differential equations of the modelling framework in their weak and strong forms. To this end, it is convenient to define the elastic strain energy density, considering the volumetric-deviatoric split considered above,
\begin{equation}\label{eq:dam-STenrgy}
\psi (\bm{\varepsilon} (\mathbf{u})) =\left( 1 - \phi \right)^2 \psi_0^+ (\bm{\varepsilon} (\mathbf{u})) + \psi_0^- (\bm{\varepsilon} (\mathbf{u})) \, .
\end{equation}

Then, in a variationally consistent fashion, the Cauchy stress is defined as,
\begin{equation}\label{eq:Cauchy}
\bm{\sigma} = \frac{\partial \psi \left( \bm{\varepsilon} (\mathbf{u}) \right) }{\partial \bm{\varepsilon} (\mathbf{u})} = \left( 1 - \phi \right)^2 \frac{\partial \psi_0^+ \left( \bm{\varepsilon} (\mathbf{u}) \right) }{\partial \bm{\varepsilon} (\mathbf{u})}+\frac{\partial \psi_0^- \left( \bm{\varepsilon} (\mathbf{u}) \right) }{\partial \bm{\varepsilon} (\mathbf{u})} = \left( 1 - \phi \right)^2 \bm{\sigma}_0^+ + \bm{\sigma}_0^-,
\end{equation}

\noindent where $\bm{\sigma}_0^+$ denotes the positive part of the undamaged Cauchy stress tensor, which undergoes degradation due to the evolution of damage, while $\bm{\sigma}_0^-$ represents the corresponding negative counterpart. Then, considering both the strain energy decomposition (\ref{eqn:str-decomposition}) and the choice of crack density function (\ref{eq:CrackDensityFunction}), the regularised functional (\ref{eq:EgriffithR}) can be expressed as,
\begin{align} 
    \mathcal{E}_\ell \left( \mathbf{u}, \phi \right) = &  \int_\Omega \left[( \left( 1 - \phi \right)^2 \psi^+_0 \left( \bm{\varepsilon} \left( \mathbf{u} \right) \right)+\psi^-_0 \left( \bm{\varepsilon} \left( \mathbf{u} \right) \right)  \right]  \, \text{d} V \nonumber \\ 
   & + \int_{V} G_c \left(\frac{1}{2\ell}\phi^2 + \frac{\ell}{2} \lvert\nabla \phi\rvert^{2}\right) \, \text{d} V   -   \int_\Omega  \mathbf{b} \cdot \mathbf{u} \, \text{d}V - \int_{\partial \Omega}  \mathbf{T} \cdot \mathbf{u} \, \text{d}S \label{eq:El2}
\end{align}

Consequently, taking the stationary of the functional (\ref{eq:El2}), using Gauss' divergence theorem and noting that the resulting expression must hold for any kinematically admissible variations of virtual quantities, the local balance equations of the problem are obtained as follows:
\begin{align}\label{eqn:strongForm}
\nabla \cdot \left[ \left( 1 - \phi \right)^2 \bm{\sigma}_0^+ + \bm{\sigma}_0^- \right]+ \boldsymbol{b} &= 0 \hspace{3mm} \text{in} \hspace{3mm} \Omega \nonumber \ \\
G_{c} \left( \dfrac{\phi}{\ell} - \ell \Delta \phi \right) - 2(1-\phi) \psi_0^+ \left( \bm{\varepsilon} \left( \mathbf{u} \right) \right) &= 0 \hspace{3mm} \text{in} \hspace{3mm} \Omega
\end{align}

Finally, one should note that crack healing is possible in the absence of supplementary constraints in the phase field evolution equation (\ref{eqn:strongForm})b. To preclude this phenomenon, a history field $\mathcal{H}$ can be introduced, $\mathcal{H}=\max _{\tau \in[0, t]} \psi_0^+(\tau)$, replacing $\psi_0^+$ as the fracture driving force \cite{Miehe2010a}. Also, a residual stiffness can be added to the degradation function to prevent ill-conditioning in fully damaged regions, such that $g(\phi)=(1-\phi)^2+\kappa$, where $\kappa$ is a small number ($\kappa=1\times10^{-7}$). 

\subsubsection{Numerical implementation}
\label{Sec:NumImplementation}

The coupled system of equations is solved using the finite element method. As described in \ref{App:FEM}, the components of the stiffness matrix and the residuals can be obtained from Eq. (\ref{eq:El2}) using the finite element discretisation. Here, the focus is on simplicity and accordingly, the model is implemented in the commercial finite element package \texttt{Abaqus} without the need for an element-level implementation. As shown in \ref{Sec:HeatApp}, the phase field evolution equation, Eq. (\ref{eqn:strongForm}b), takes the form of Poisson's equation, such that one can exploit the analogy with the steady-state heat transfer equation and use in-built \texttt{Abaqus} capabilities. By treating the phase field variable $\phi$ akin to the temperature field and applying the crack driving force via an appropriate (nonlinear) heat source, one can easily implement the phase field model at the integration point level, using a user material (UMAT) subroutine. Details of the implementation are provided in \ref{Sec:HeatApp} and Refs. \cite{AES2021,Materials2021}. A monolithic solution scheme was used to solve the coupled displacement and phase field equations, ensuring unconditional stability and thus maximising efficiency. 

\subsection{Numerical experiments: defining the boundary value problem}

\subsubsection{Challenge data, requisitions and characteristics}

\begin{figure}[H]
    \centering
    \begin{subfigure}[t]{0.45\textwidth}
    \includegraphics[width=\textwidth]{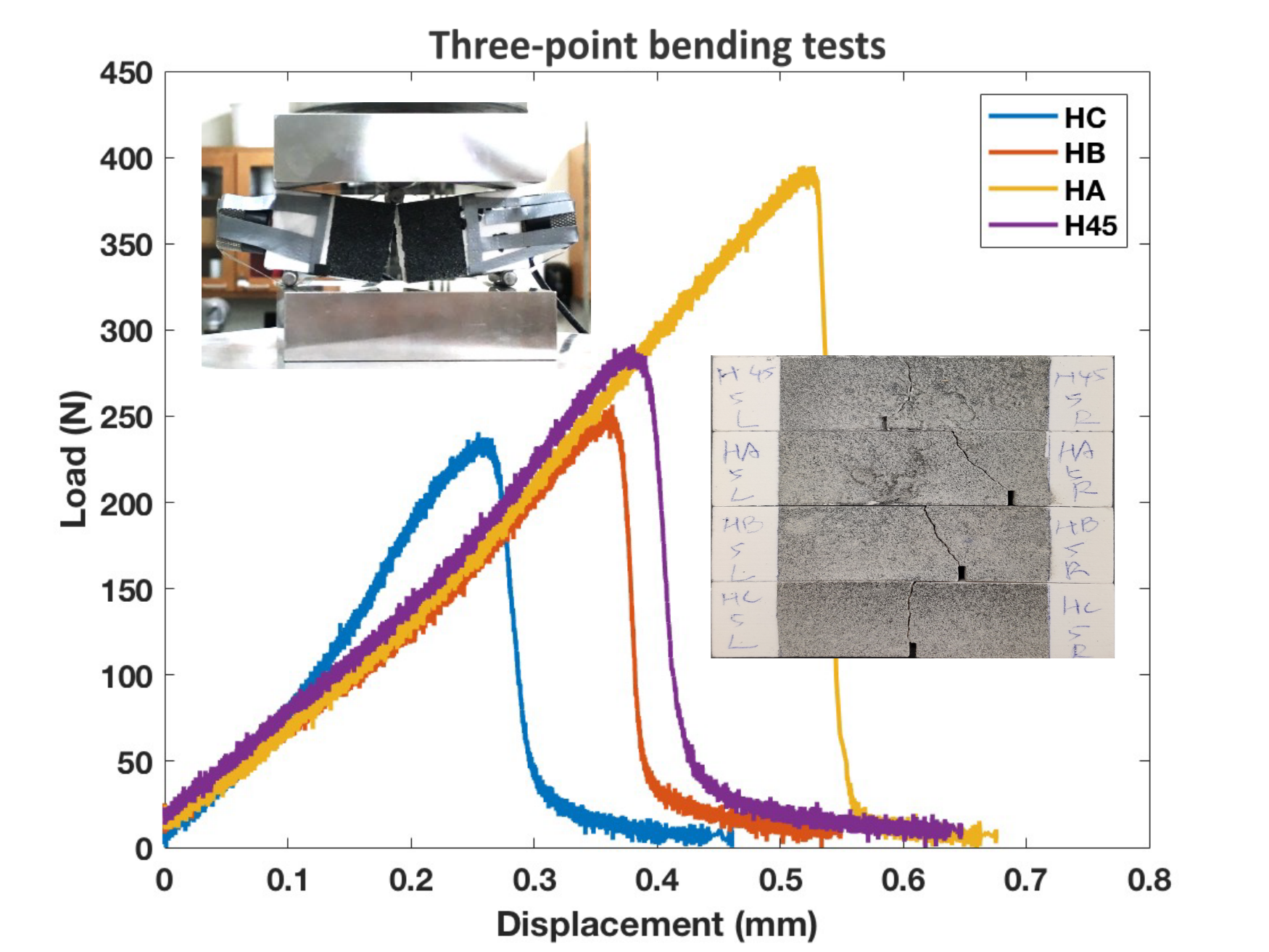}
    \caption{}
    \label{fig:Experiments-a}
    \end{subfigure}
    \begin{subfigure}[t]{0.45\textwidth}
    \includegraphics[width=\textwidth]{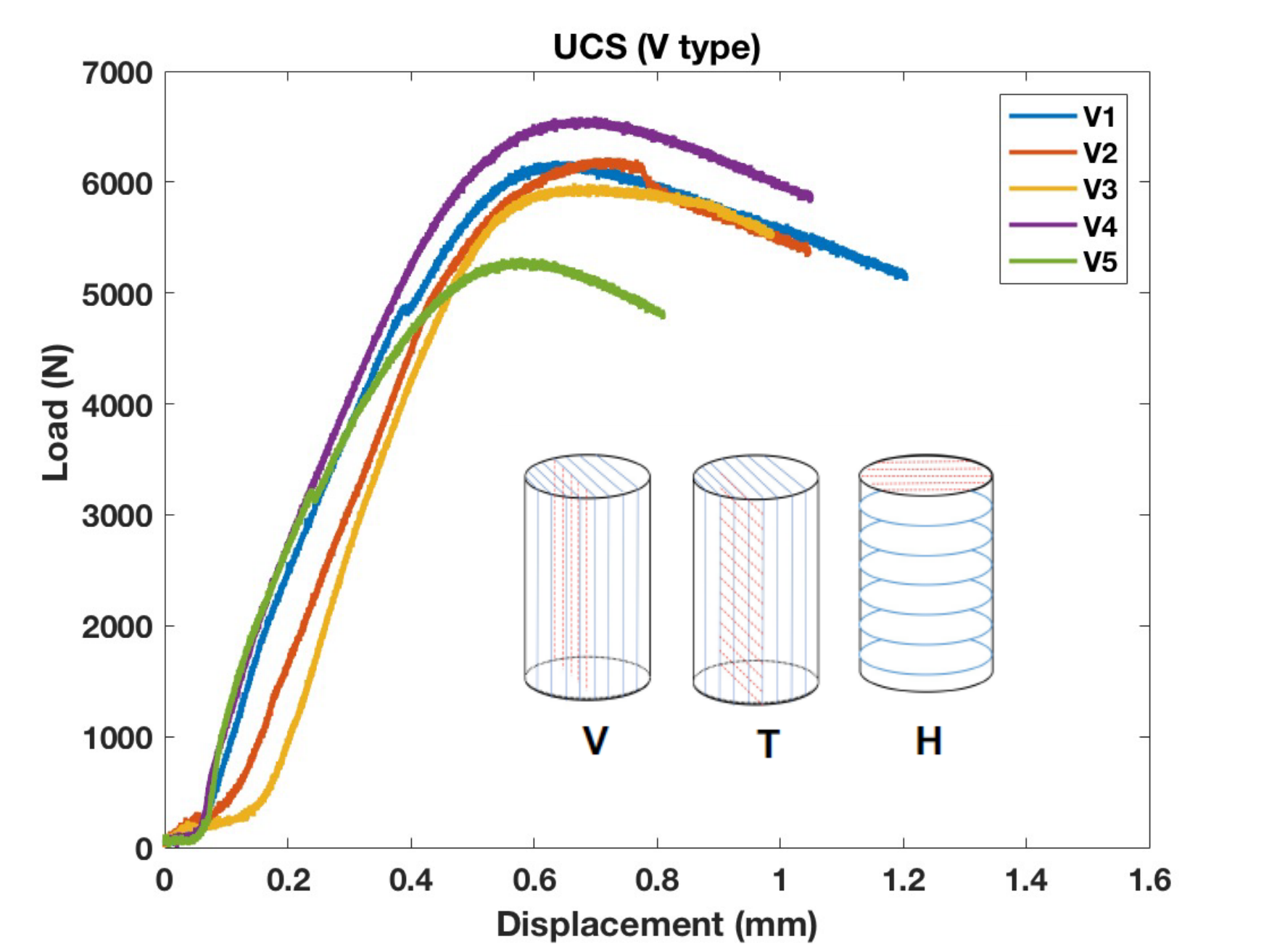}
    \caption{}
    \label{fig:Experiments-b}
    \end{subfigure}
    \begin{subfigure}[t]{0.45\textwidth}
    \includegraphics[width=\textwidth]{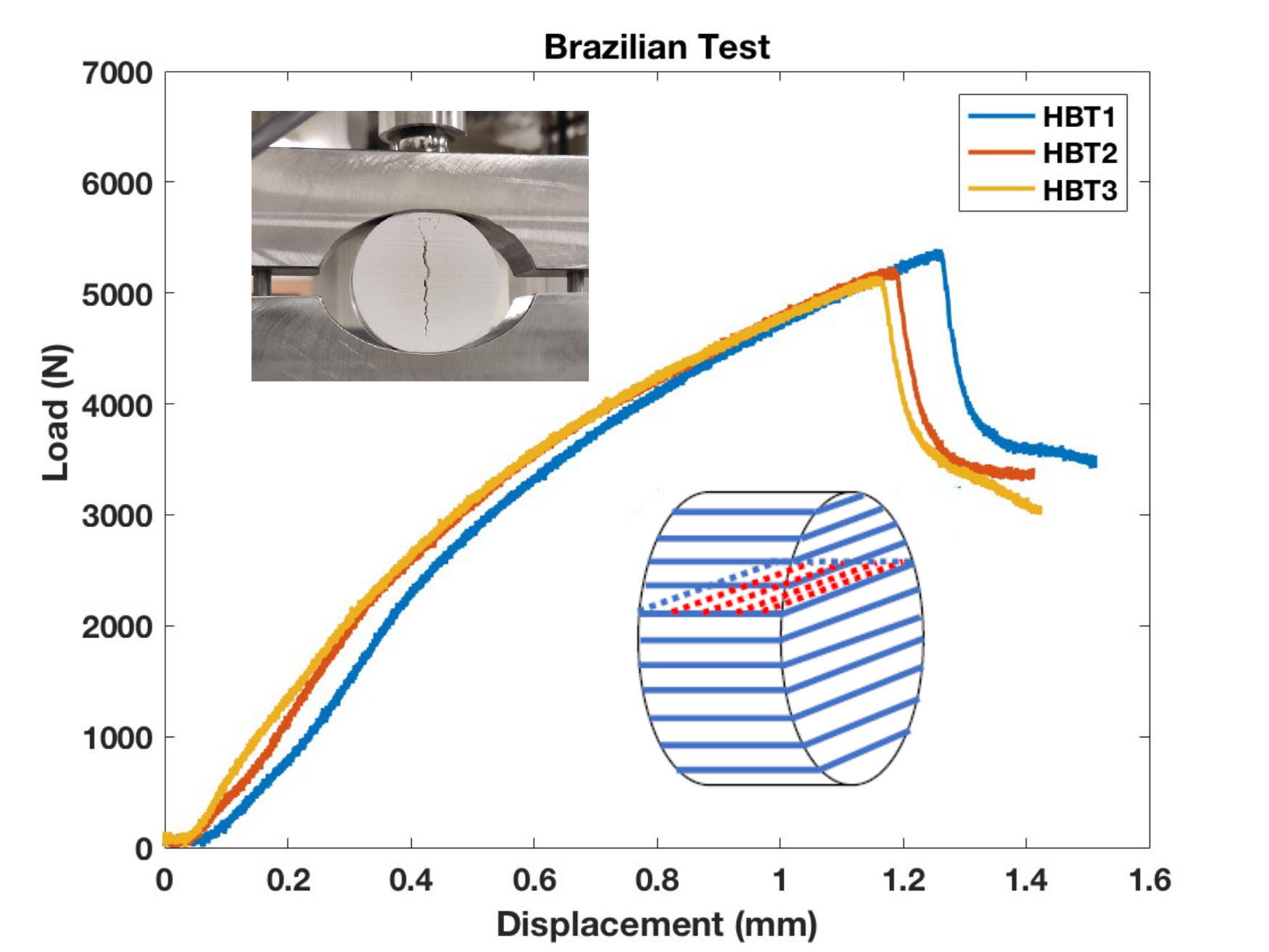}
    \caption{}
    \label{fig:Experiments-c}
    \end{subfigure}
    \begin{subfigure}[t]{0.45\textwidth}
    \includegraphics[width=\textwidth]{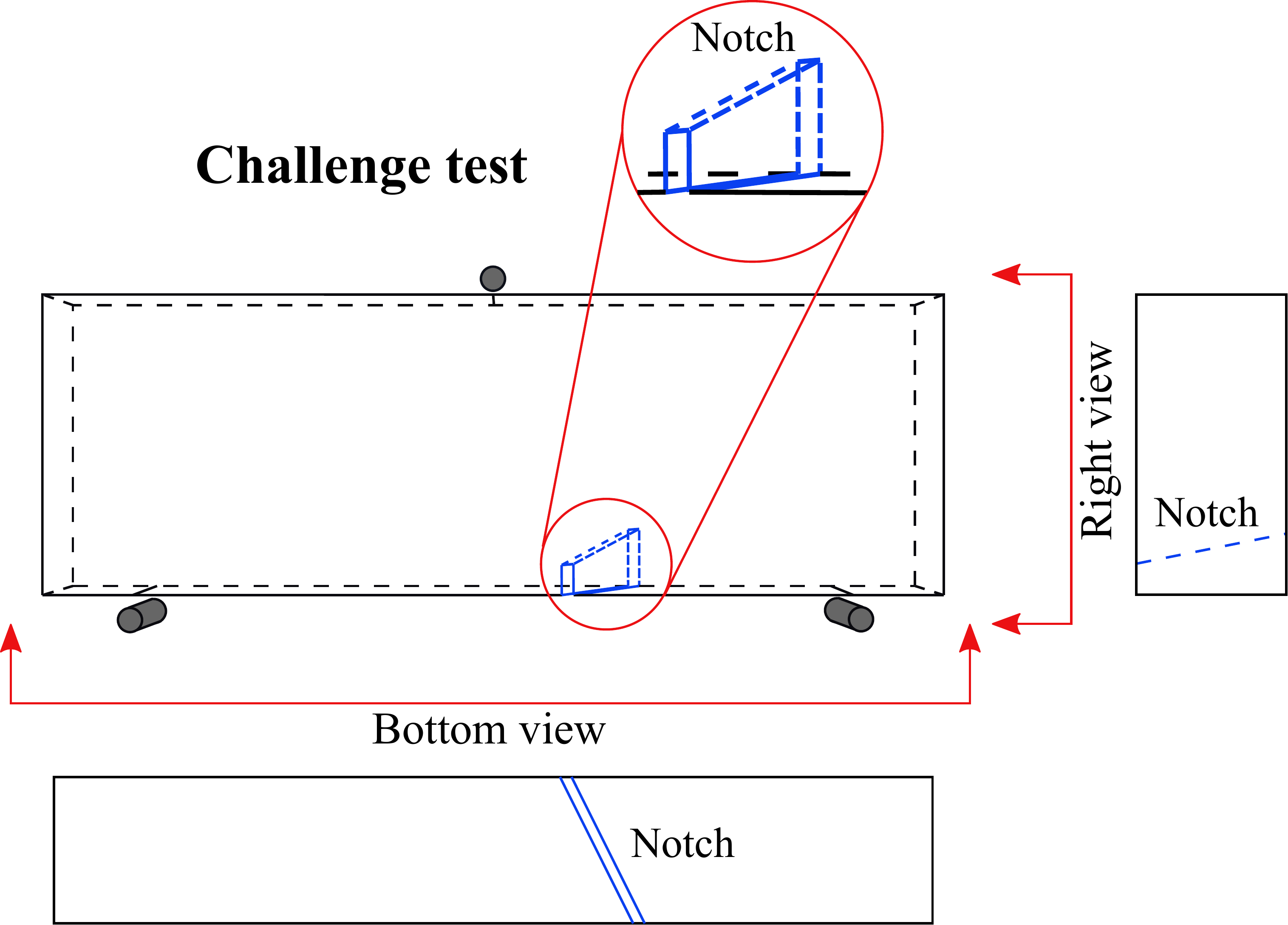}
    \caption{}
    \label{fig:Experiments-d}
    \end{subfigure}
    \caption{Damage Mechanics Challenge data: calibration and to-be-predicted experiments. The calibration data included force versus displacement measurements for: (a) four three-point bending configurations (of equal dimensions but different notch configurations), (b) unconfined compressive tests, and (c) Brazilian disk tests. Based on this information, participants were asked to blindly estimate the cracking characteristics (peak load, crack trajectory and morphology) of an unconventional three-point bending test with an inclined, unsymmetric notch (d).}
    \label{fig:Experiments}
\end{figure}

The phase field fracture model was used to conduct numerical experiments with the aim of benchmarking model predictions against calibration data and providing a blind estimate of the required outputs. The samples needed for the challenge experiment and the calibration data were manufactured by the hosts of the \emph{Damage Mechanics Challenge} using additive manufacturing. The material employed was a special type of gypsum, which resulted from the bonding of calcium sulfate hemihydrate layers (bassanite powders with a deposition layer thickness of 0.1 mm) with a proprietary water-based binder (ProJet X60 VisiJet PXL). For consistency, all samples were obtained from the same 3D printing build. As summarised in Fig. \ref{fig:Experiments}, the calibration data included: (i) three-point bending tests on four types of notched samples, differentiated by the location of their notch, (ii) uniaxial loading tests to measure the unconfined compressive strength (UCS), and (iii) Brazilian tests to measure the tensile strength. Based on this information, participants were provided with the geometry of the challenge test, a three-point bending test with an inclined notch - see Fig. \ref{fig:Experiments}d, and asked to numerically predict: 
\begin{itemize}
    \item The force versus displacement response.
    \item The crack trajectory to benchmark against Digital Image Correlation (DIC) images.
    \item The crack surface morphology to benchmark against laser profilometry measurements.
\end{itemize}

Because the growth of a pre-existing notch in a three-point bending test is likely to be driven by tensile stress states, we disregarded the UCS values measured in the calibration unconfined compressive tests. However, these experiments can be useful in providing a rough estimate of the material's Young's modulus $E$. Specifically, the data provided in Fig. \ref{fig:Experiments}c can be well-fitted with $E$ values between 900 and 1100 MPa. Nevertheless, it is important to emphasise that this is likely to be a higher value than that relevant to tensile loading, due to the additional stiffness provided by existing defects under compression. The Brazilian test data provided was also disregarded. While the Brazilian test is an experiment frequently used to estimate a material's tensile strength, the analysis of the experiment with the \texttt{BrazVal} App \cite{IJRMMS2022} revealed that the conditions of validity of the test were not fulfilled. As elaborated in Ref. \cite{IJRMMS2022}, jaws with sufficiently small radii must be used to ensure that cracking initiates in the disk centre. If this is not the case, the tensile strength obtained from the peak load measurement is an underestimation of the real material tensile strength. As a result, the tensile strength corresponding to the average peak load in the Brazilian tests conducted (5200 N, $\sigma_c \approx 2.5$ MPa) is deemed to be an unsuitable value and can only serve as a lower bound. Accordingly, the calibration three-point bending tests were used to estimate the three parameters of our phase field fracture model: Young's modulus $E$, strength $\sigma_c$ and toughness $G_c$. Poisson's ratio was assumed to be equal to $\nu=0.2$ as is commonly the case in rock-like materials and calculations with other values (results not shown here) indicate negligible differences. While a mode I fracture experiment can serve to independently calibrate each of these parameters, it is important to emphasise that they all have a physical meaning and can be independently measured. \\

\begin{figure}[H]
    \centering
    \begin{subfigure}[t]{0.405\textwidth}
    \includegraphics[width=\textwidth]{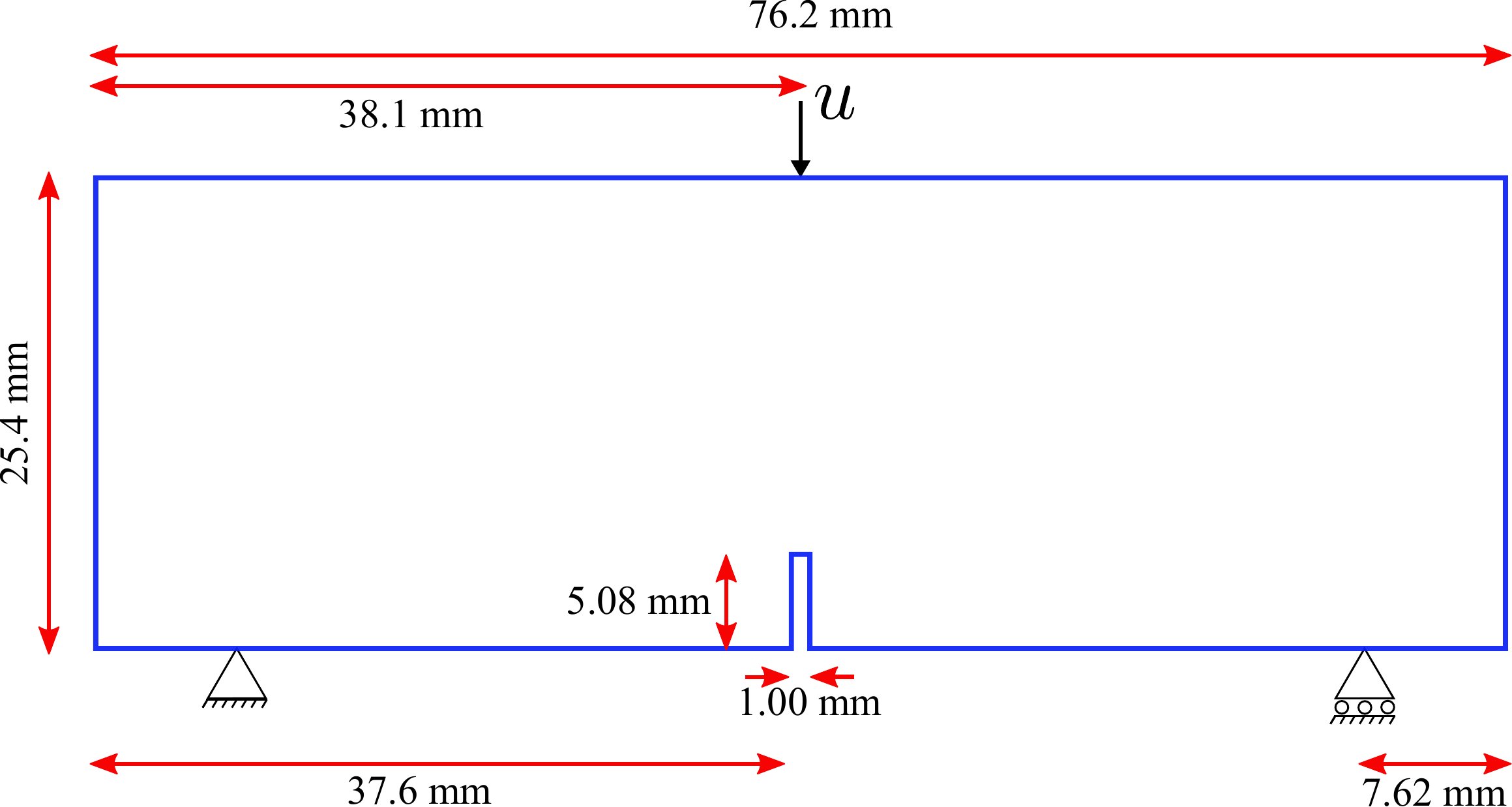}
    \caption{}
    \label{}
    \end{subfigure}\hspace{0.09\textwidth}
    \begin{subfigure}[t]{0.4\textwidth}
    \includegraphics[width=\textwidth]{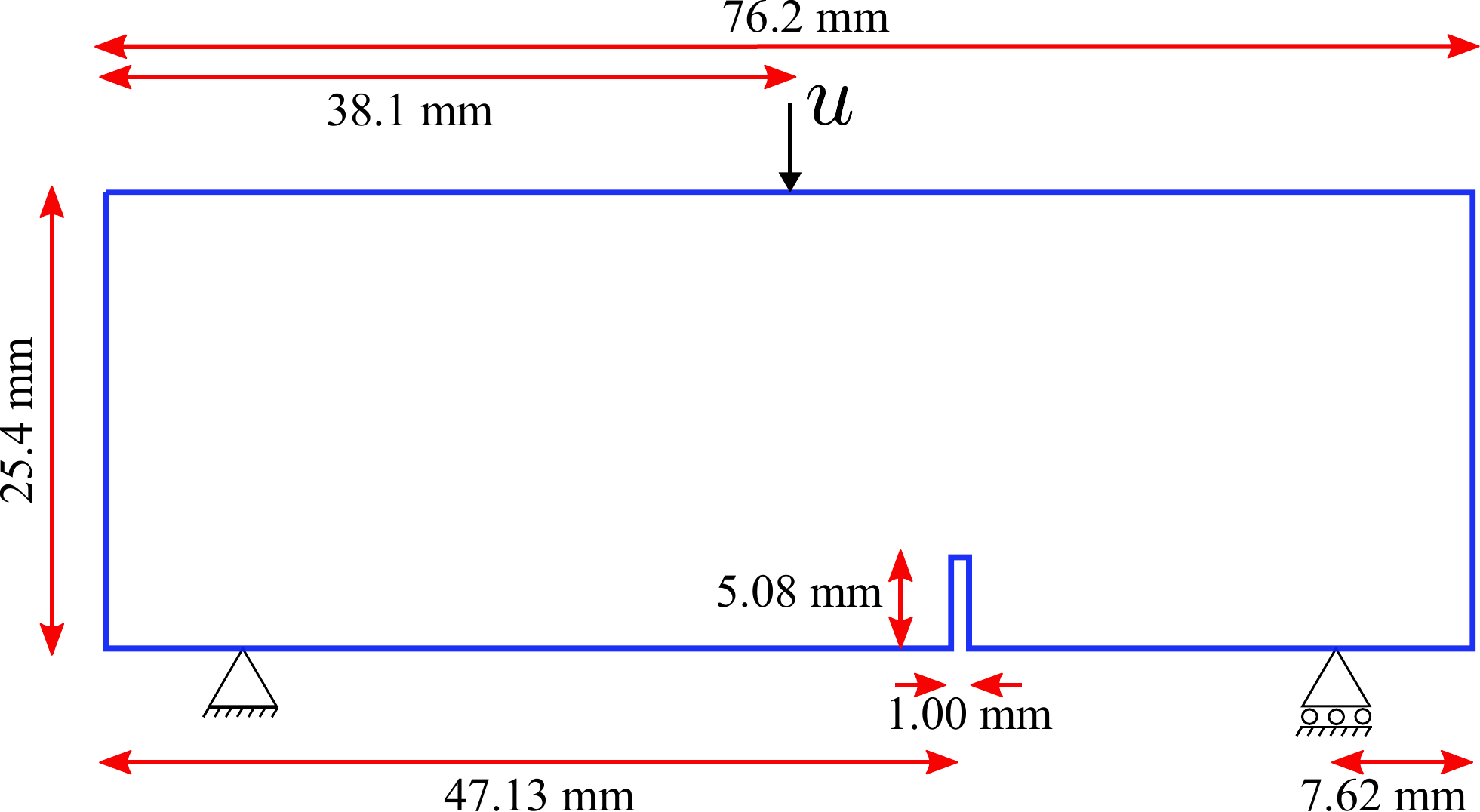}
    \caption{}
    \label{}
    \end{subfigure}
    \begin{subfigure}[t]{0.4\textwidth}
    \includegraphics[width=\textwidth]{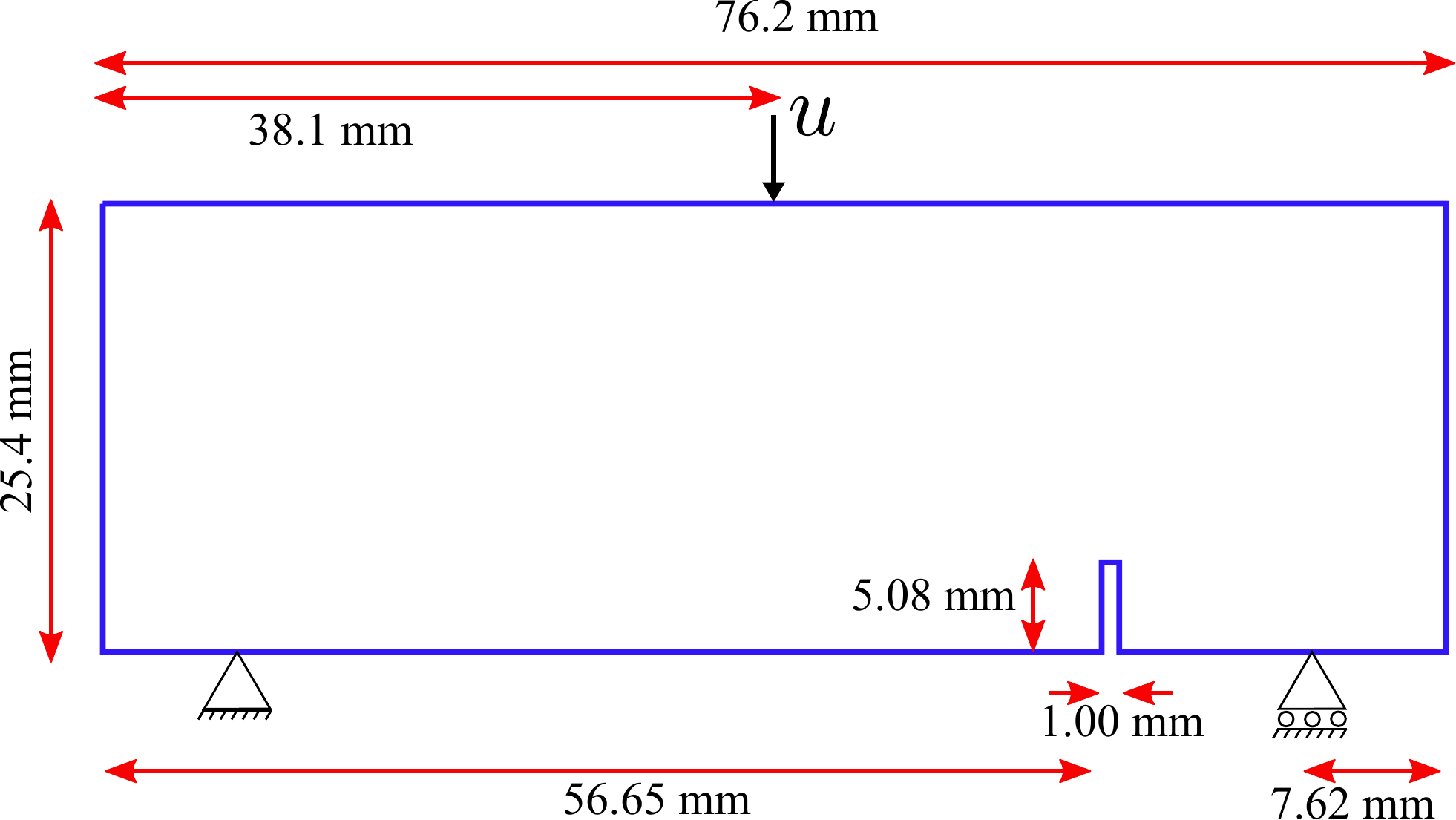}
    \caption{}
    \label{}
    \end{subfigure}\hspace{0.09\textwidth}
    \begin{subfigure}[t]{0.35\textwidth}
    \includegraphics[width=\textwidth]{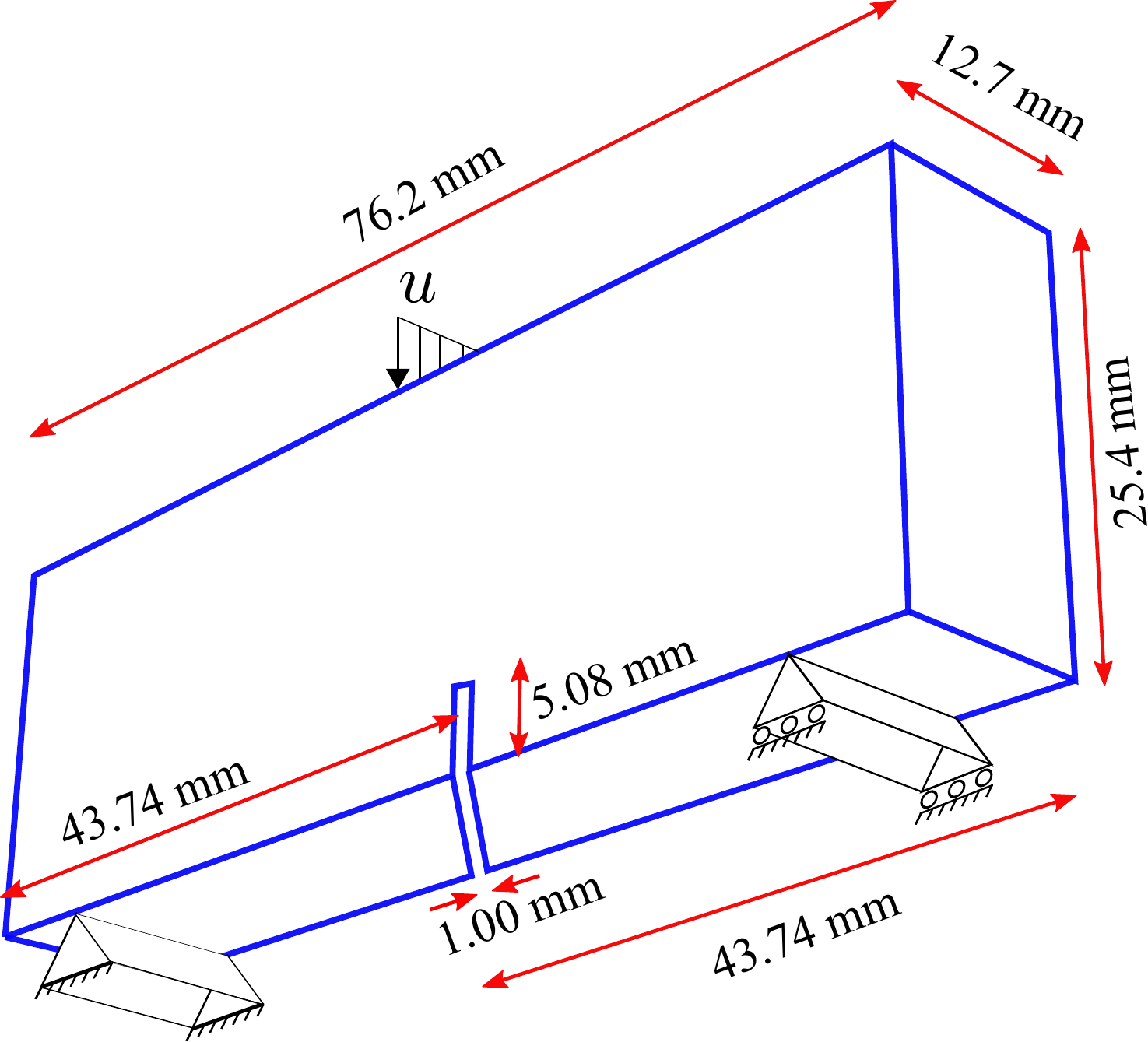}
    \caption{}
    \label{}
    \end{subfigure}
    \caption{Three-point bending tests used for generating calibration data: geometry, dimensions and boundary conditions. The tests included a conventional three-point bending experiment, denoted as HC (a), two tests where the notch was placed eccentric, HB (b) and HA (c), and a fourth experiment where the notch was inclined 45$^\circ$ along the thickness, requiring a 3D analysis, H45 (d).}
    \label{fig:SketchSamples}
\end{figure}

Three-point bending tests were conducted on five types of samples, with the results obtained for four of them provided as calibration data. These samples are illustrated in Fig. \ref{fig:SketchSamples}, providing details of their geometry and boundary conditions. All the samples had the following dimensions: 25.4 mm $\times$ 76.2 mm $\times$ 12.7 mm. The tests denoted HC correspond to a standard three-point bending configuration, with the notch located in the centre of the sample and aligned with the applied load. In tests HB and HA the notch was placed eccentric to break the symmetry of the beam and induce mixed-mode fracture conditions. A fourth calibration test, denoted H45, included a notch inclined 45$^\circ$ along the thickness, requiring full 3D analysis. On the other hand, the challenge experiment was based on a more intricate geometry, containing a crack that was not only inclined along the out-of-plane direction but also exhibited a variation in notch depth along the sample thickness - see Figs. \ref{fig:Experiments}d and \ref{fig:Challenge-Configll}. An important point to emphasise is that, as will be shown below, the mode I HC three-point bending experiment suffices to estimate the parameters of the phase field fracture model; the model can predict (without any additional fitting) the mixed mode behaviour of the remaining three-point bending as crack trajectories are naturally captured following the path of maximum energy release rate.

\subsubsection{Computational details}

\begin{figure}[H]
    \centering
    \begin{subfigure}[t]{0.4\textwidth}
    \includegraphics[width=\textwidth]{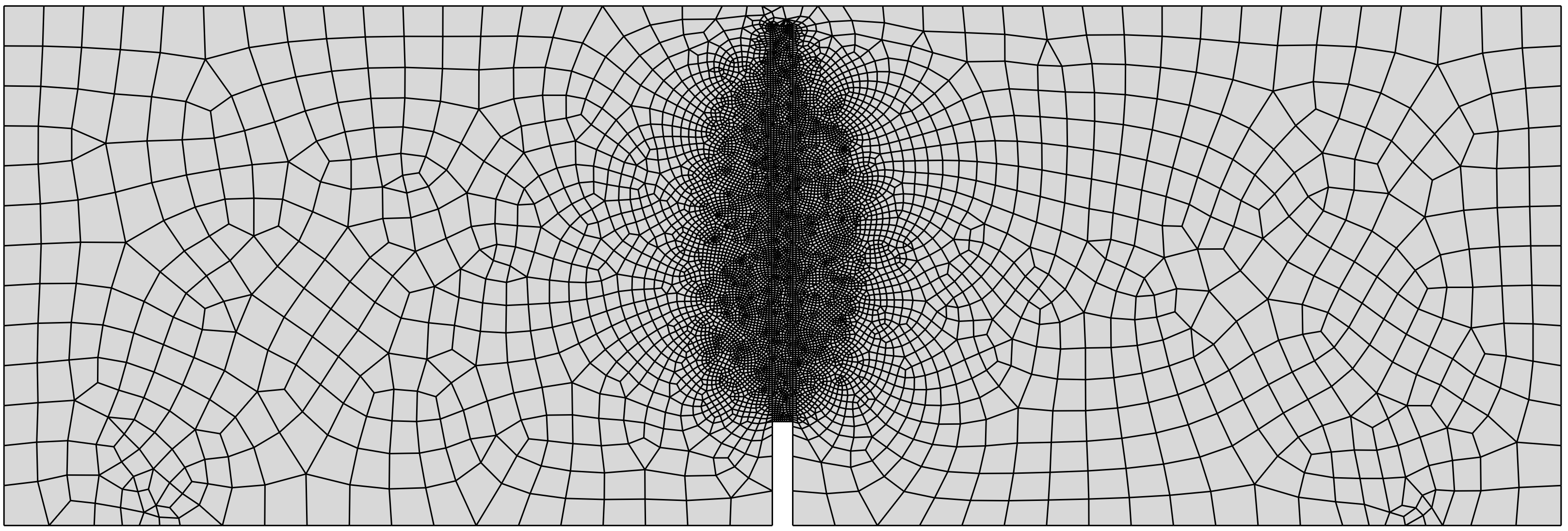}
    \caption{}
    \label{}
    \end{subfigure}\hspace{0.09\textwidth}
    \begin{subfigure}[t]{0.4\textwidth}
    \includegraphics[width=\textwidth]{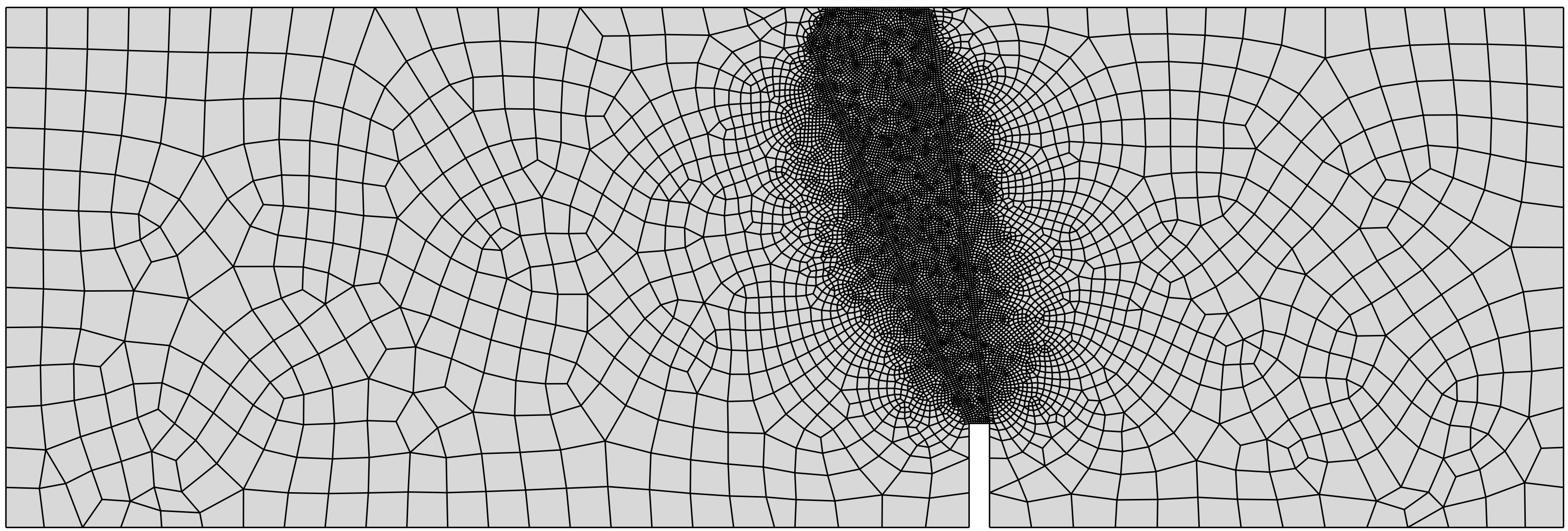}
    \caption{}
    \label{}
    \end{subfigure}
    \begin{subfigure}[t]{0.4\textwidth}
    \includegraphics[width=\textwidth]{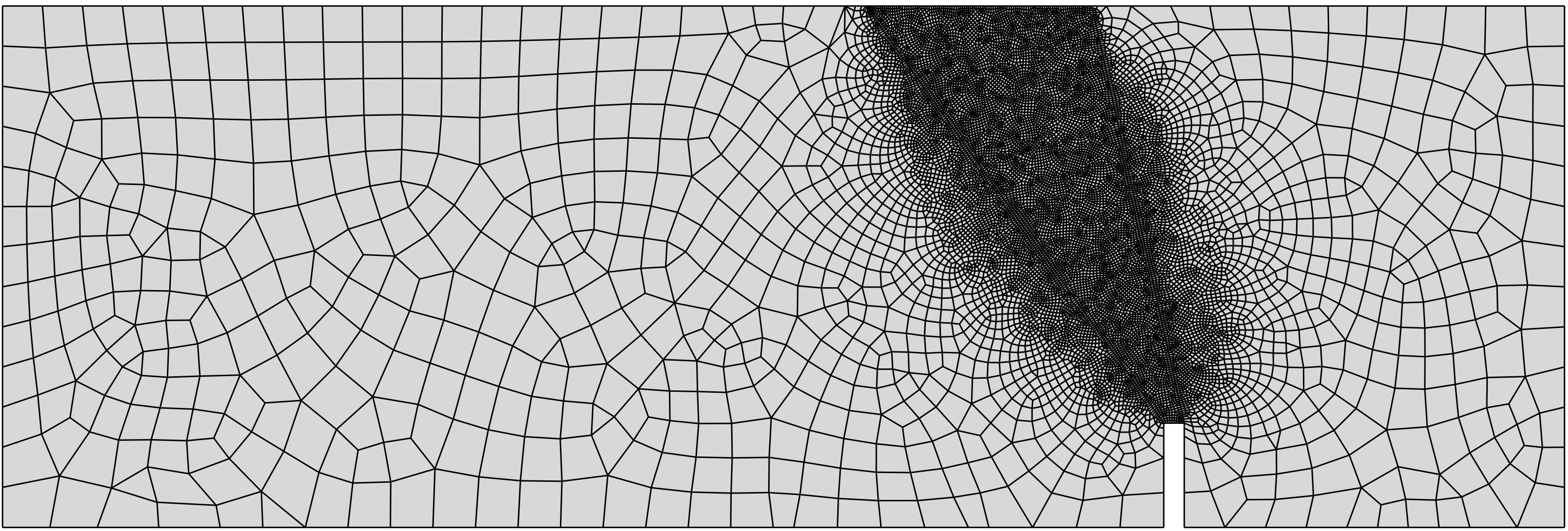}
    \caption{}
    \label{}
    \end{subfigure}\hspace{0.09\textwidth}
    \begin{subfigure}[t]{0.35\textwidth}
    \includegraphics[width=\textwidth]{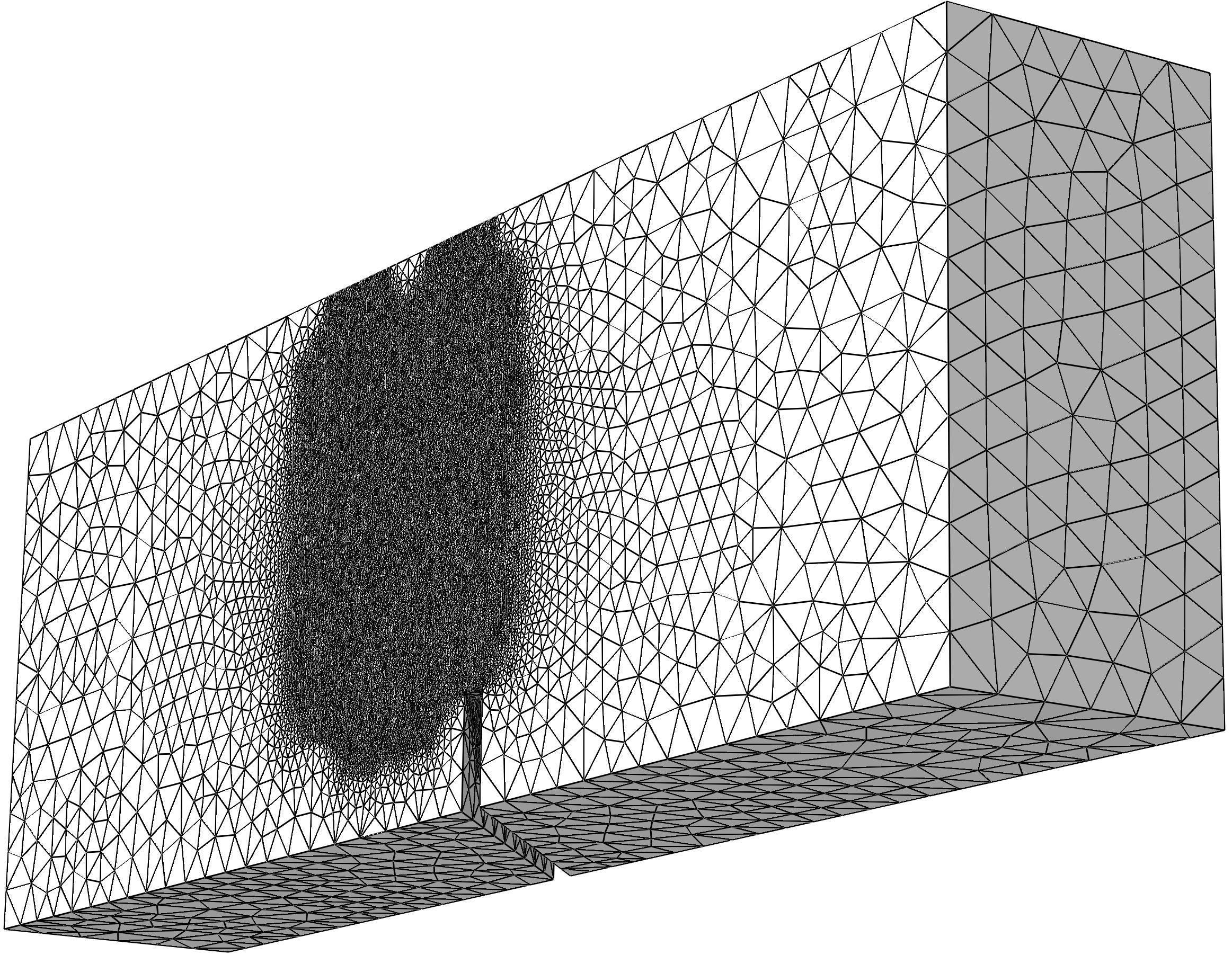}
    \caption{}
    \label{}
    \end{subfigure}
    \caption{Finite element discretisation of the three-point bending tests used for generating calibration data. The model HC (a) employs a total of 8,888 bi-linear quadrilateral elements, the model HB (b) uses 11,242 bi-linear quadrilateral elements, the model HA (c) uses 14,622 bi-linear quadrilateral elements, and the three-dimensional model H45 employs 1,953,053 linear tetrahedral elements. Since the crack trajectory is not known a priori, the mesh is strategically refined in regions of potential crack growth.}
    \label{fig:Calib_Geometry}
\end{figure}

The finite element meshes employed for each of the calibration three-point bending tests are given in Fig. \ref{fig:Calib_Geometry}. Plane strain conditions were assumed for calibration tests HA, HB and HC, whereas a 3D model has to be employed for H45. Three and four degrees of freedom per node are respectively employed in the 2D and 3D models, involving the components of the displacement vector and the scalar phase field variable. A key computational advantage of the phase field fracture model is its ability to deliver mesh-independent results, due to its non-local nature. However, this requires a mesh sufficiently fine to resolve the phase field length scale $\ell$. Specifically, it has been shown that the characteristic element size has to be five times smaller than $\ell$ to ensure mesh objectivity \cite{PTRSA2021}. This rule is followed in all our calculations, requiring the use of a refined mesh along the potential crack propagation region. Since the crack path is not known a priori, the mesh is refined over a relevant, sufficiently large region near the notch - see Fig. \ref{fig:Calib_Geometry}. 

\begin{figure}[H]
    \centering
    \begin{subfigure}[t]{0.4\textwidth}
    \includegraphics[width=\textwidth]{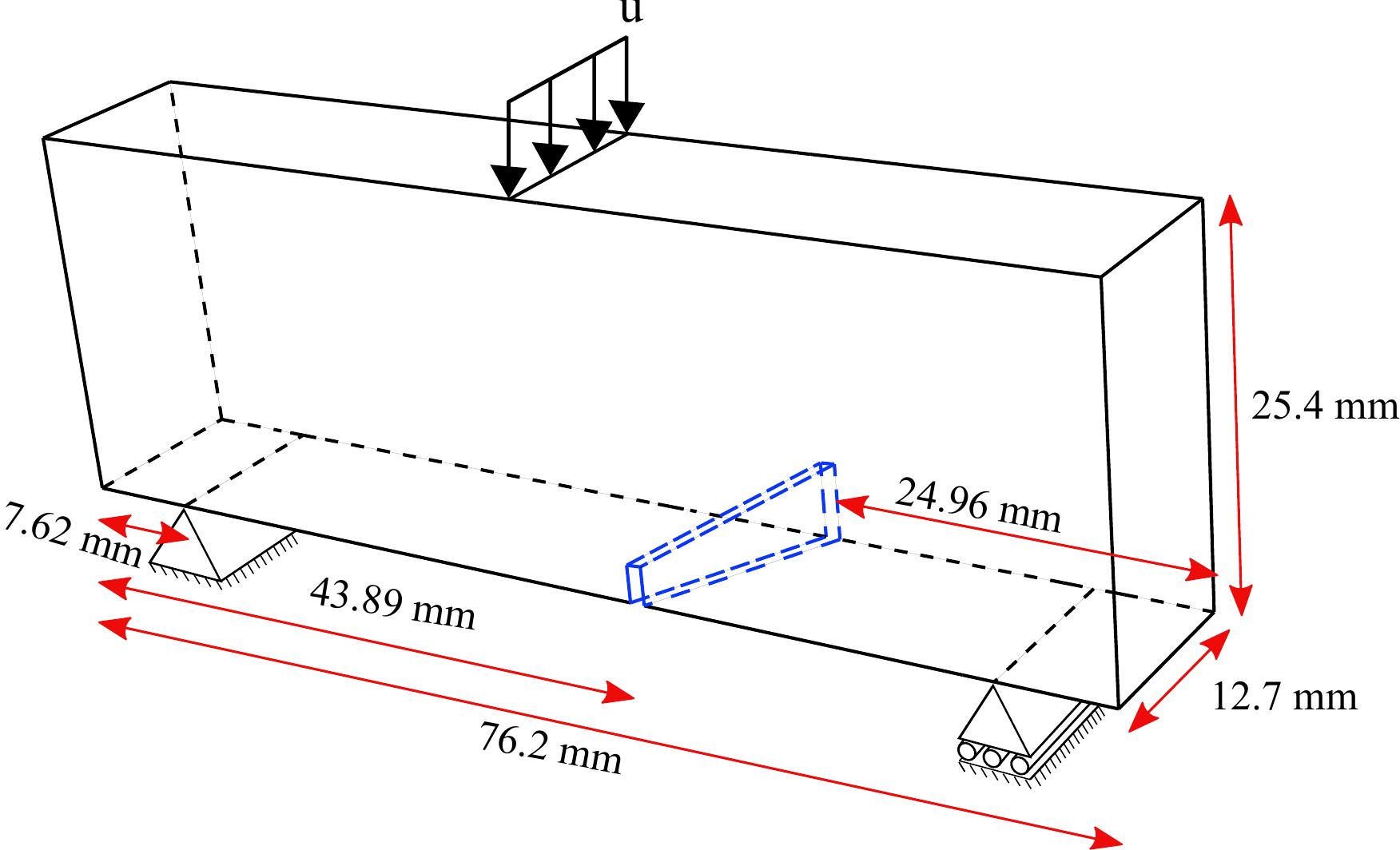}
    \caption{}
    \label{}
    \end{subfigure}
    \begin{subfigure}[t]{0.4\textwidth}
    \includegraphics[width=\textwidth]{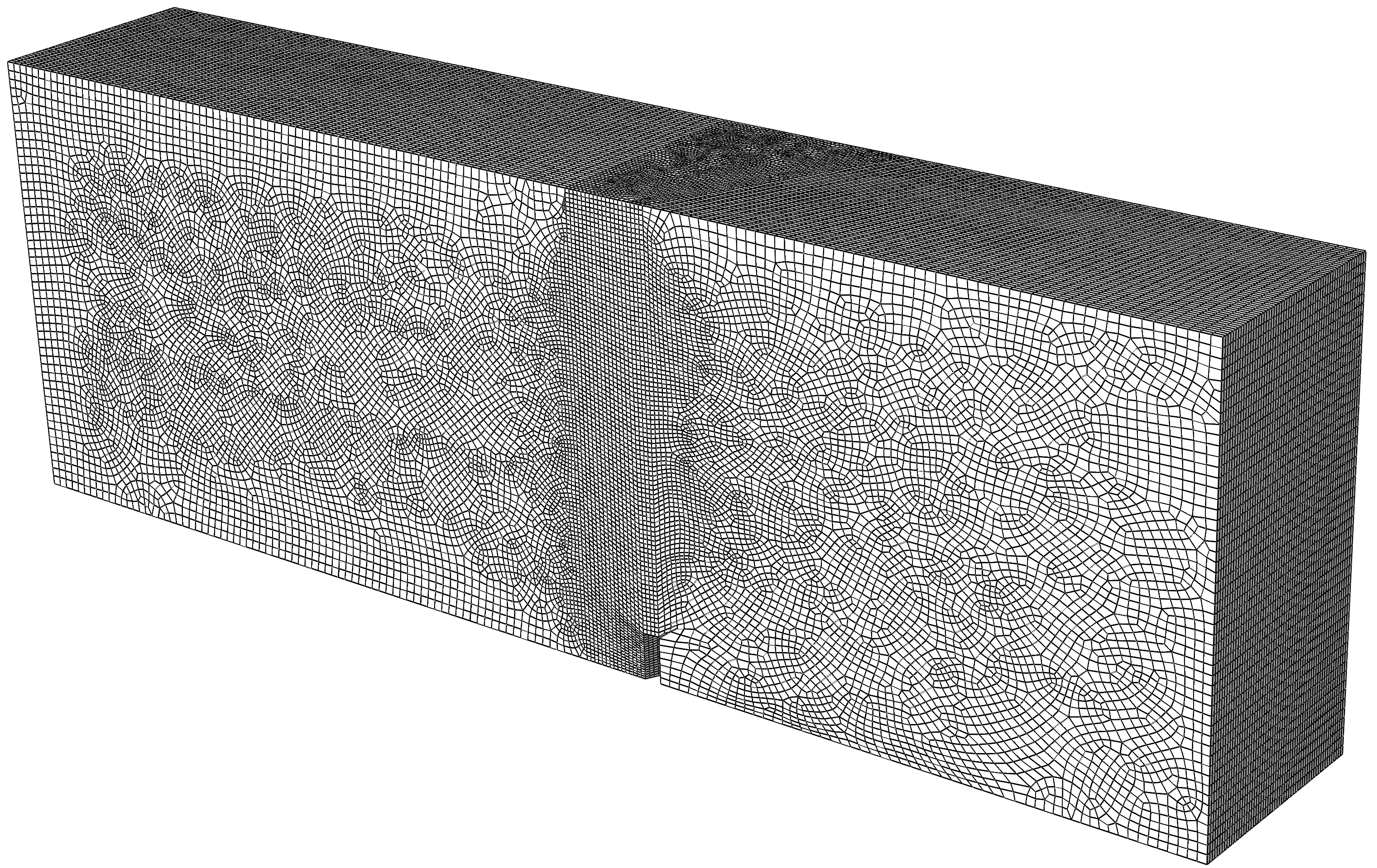}
    \caption{}
    \label{}
    \end{subfigure}
    \caption{Details of the challenge test: (a) Geometry, dimensions and boundary conditions, and (b) finite element discretisation, employing a total of 904,429 linear brick elements.}
    \label{fig:Challenge-Configll}
\end{figure}

Four-node quadrilateral elements with full integration are employed for the 2D case studies, while four-node linear tetrahedral elements are used for the 3D benchmark. For the sake of comparison, both tetrahedral and brick elements are used in the predictions of the challenge test, as discussed below. As quantified in the caption of Fig. \ref{fig:Calib_Geometry}, the finite element meshes range from 8,000 to 15,000 elements, in the 2D cases, while close to 2 million elements are employed for the 3D analysis. Calculation times go from 30 minutes for the 2D analyses (using a single core) to eight days for the 3D model (using 8 cores). A direct linear solver is employed. In the case of the challenge test, the total number of DOFs was close to 4 million, and calculations exceeded ten days on an Intel(R) Xeon(R) Gold 6242R workstation using 16 cores. These calculation times are intrinsically related to the choice of a fully monolithic scheme, which provides accuracy and unconditional stability at the expense of poor convergence. Calculation times can be very significantly reduced through the use of staggered or BFGS-based monolithic approaches. The geometry, mesh and boundary conditions of the challenge test are provided in Fig. \ref{fig:Challenge-Configll}. 

\subsubsection{Calibration of the material parameters}

Preliminary calculations were conducted to estimate the values of the three input parameters of the model: Young's modulus $E$, toughness $G_c$ and strength $\sigma_c$, with the last one being fixed by an appropriate choice of the phase field length scale $\ell$. As discussed above, the mode I three-point bending test (denoted HC) suffices to calibrate these variables, and the quality of the calibration was then benchmarked by predicting the failure of the other three three-point bending tests, which are more intricate and inherently mixed-mode. The calibrated and verified model was then used to deliver a blind estimate for the test challenge.\\

\begin{figure}[H]
    \centering
    \begin{subfigure}[t]{0.4\textwidth}
    \includegraphics[width=\textwidth]{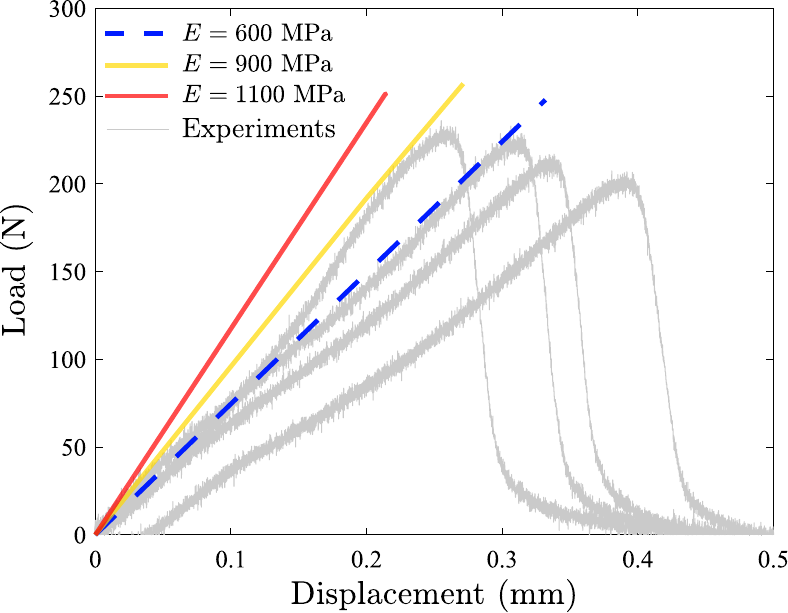}
    \caption{}
    \label{fig:Calib_Material-a}
    \end{subfigure}\hspace{0.05\textwidth}
    \begin{subfigure}[t]{0.4\textwidth}
    \includegraphics[width=\textwidth]{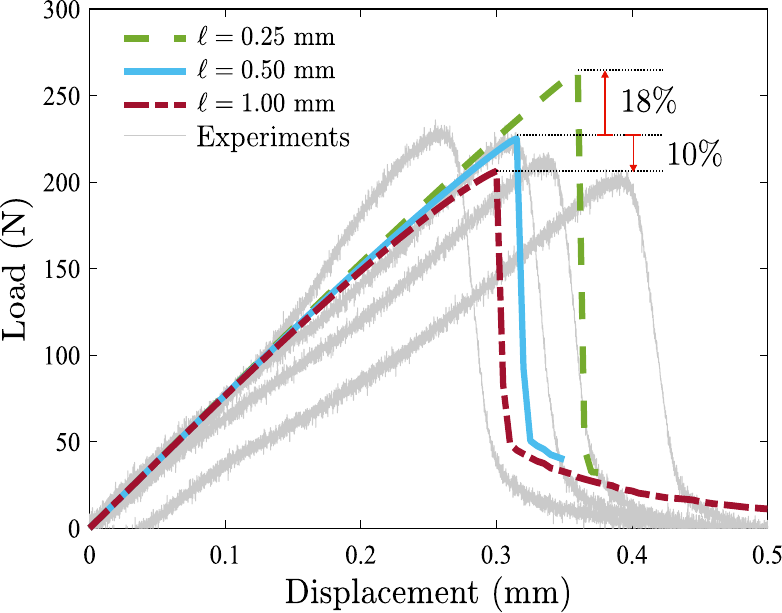}
    \caption{}
    \label{fig:Calib_Material-b}
    \end{subfigure}\hspace{0.05\textwidth}
    \begin{subfigure}[t]{0.4\textwidth}
    \includegraphics[width=\textwidth]{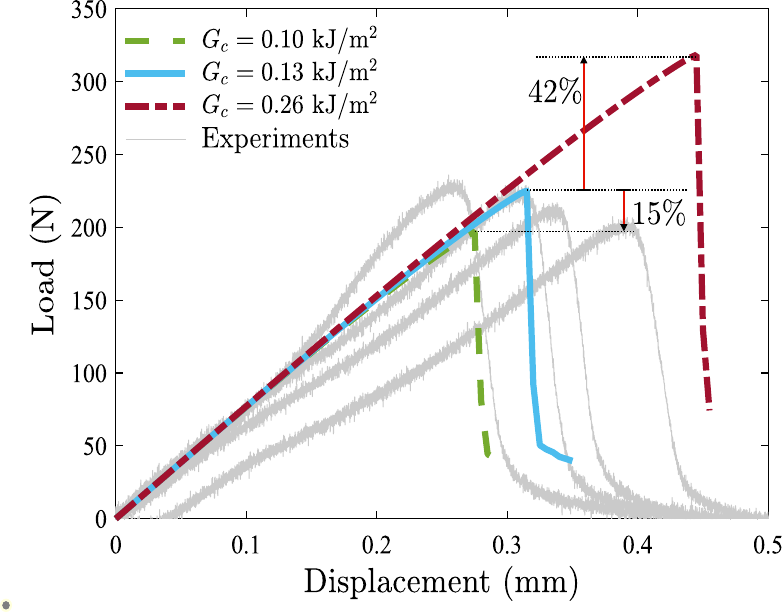}
    \caption{}
    \label{fig:Calib_Material-c}
    \end{subfigure}\hspace{0.05\textwidth}
    \begin{subfigure}[t]{0.4\textwidth}
    \includegraphics[width=\textwidth]{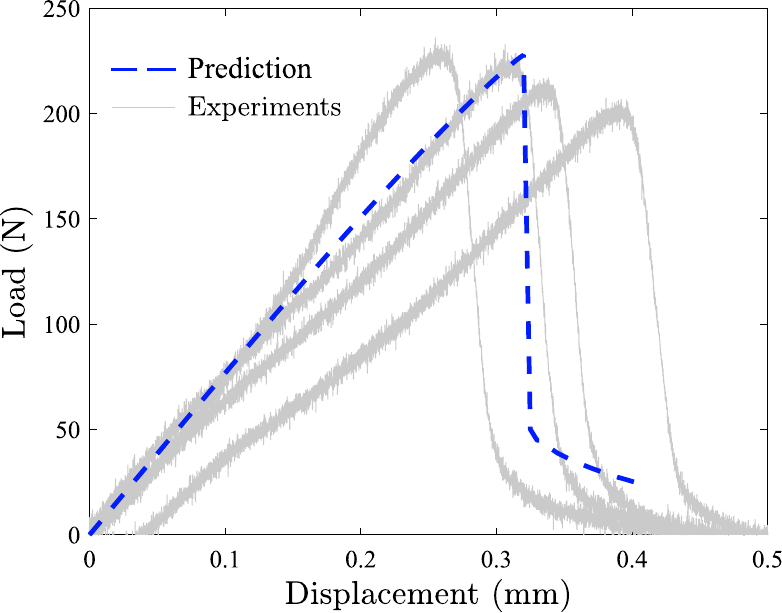}
    \caption{}
    \label{fig:Calib_Material-d}
    \end{subfigure}
    \caption{Using the mode I three-point bending experiment (HC) to estimate the three input parameters to the phase field fracture model: Young's modulus $E$, toughness $G_c$ and phase field length scale $\ell$ (or strength $\sigma_c$). Numerical force versus displacement results, and comparison with experiments, for varying (a) $E$, (b) $\ell$, and (c) $G_c$. Lastly, (d) shows the prediction obtained with the choices $E=600$ MPa, $\nu=0.2$, $G_c=0.13$ kJ/m$^2$ and $\ell=0.5$ mm. These choices give a strength of $\sigma_c=4.05$ MPa.}
    \label{fig:Calib_Material}
\end{figure}

Let us consider first the case of Young's modulus $E$. As shown in Fig. \ref{fig:Calib_Material}a, the numerical results obtained show that the range of values inferred from the UCS experiments (900-1100 MPa) overestimates the stiffness shown in the three-point bending tests. Instead, a value of $E=600$ MPa appears to provide a much better agreement. As discussed above, these discrepancies between the stiffness of tensile and compressive tests are likely to be related to the influence of defect dilation in the former (as opposed to defect closure and friction in the latter). We proceed then to determine the fracture parameters, $G_c$ and $\sigma_c$ (or $\ell$). Recall that the choice of $\ell$ defines the strength, as per Eq. (\ref{eq:Hom-Solution}). A higher sensitivity to the choice of $G_c$ is expected because the samples contain large pre-existing defects and thus failures are likely to be toughness-controlled (as opposed to strength-controlled) \cite{PTRSA2021}. This is shown in Figs. \ref{fig:Calib_Material}b and \ref{fig:Calib_Material}c; doubling the value of $G_c$ brings an increase of 46\% in the critical load, while the critical load only changes by 10\% when the value of $\ell$ is halved. But qualitatively, changes in $G_c$ and $\ell$ have the same effect - changing the magnitude of the critical load. Accordingly, we make a judicious choice and pick a pair of $G_c$ and $\ell$ values that provide a good agreement with the experimentally measured peak load while falling within the range of expected values for rock-like materials. These are $G_c=0.13$ kJ/m$^2$ and $\sigma_c=4.05$ MPa ($\ell=0.5$ mm).
As discussed below, this choice of parameters gives a remarkable agreement with the other (mixed-mode) four three-point bending tests and with the challenge test, in terms of peak load, crack trajectory and crack morphology. 

\section{Results}

We proceed to showcase the numerical results obtained using the model and boundary value problems described in Section \ref{Sec:2approach}. First, from a set of parameters calibrated with the mode I three-point bending test, we examine the ability of the model to predict cracking in the four mixed-mode three-point bending data calibration experiments (Section \ref{sec:CalibrationRes}). Then, we provide blind estimates of the failure characteristics of the challenge test (Section \ref{sec:ChallengeRes}). 

\subsection{Model benchmarking against calibration data}
\label{sec:CalibrationRes}

\begin{figure}[H]
    \centering
    \begin{subfigure}[t]{0.41\textwidth}
    \includegraphics[width=\textwidth]{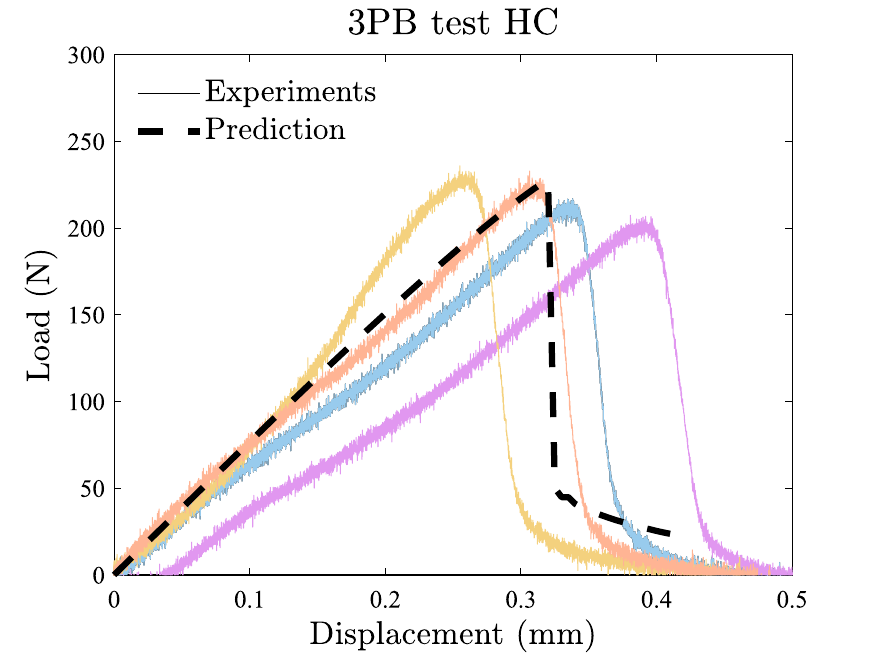}
    \caption{}
    \label{}
    \end{subfigure}\hspace{0.05\textwidth}
    \begin{subfigure}[t]{0.41\textwidth}
    \includegraphics[width=\textwidth]{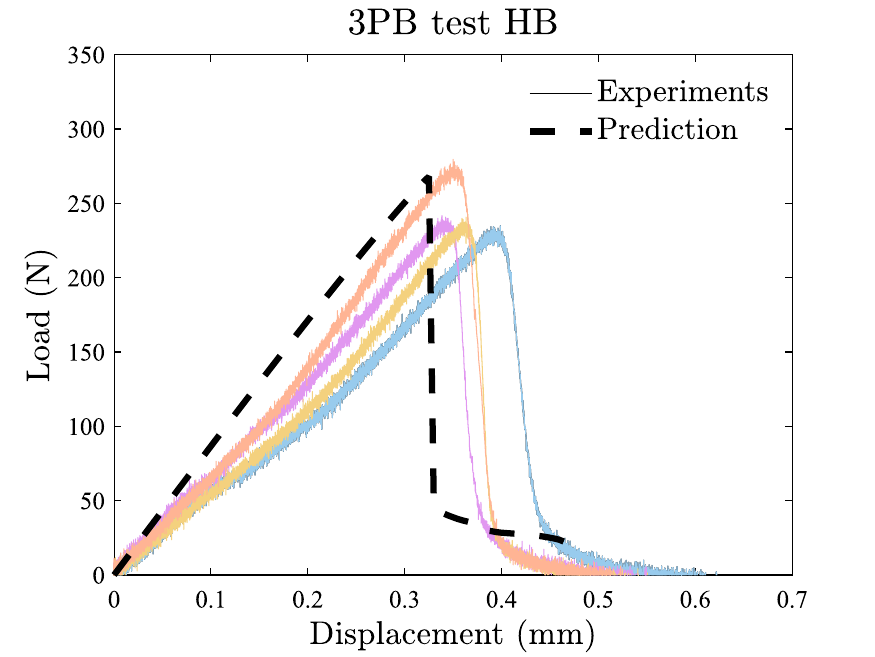}
    \caption{}
    \label{}
    \end{subfigure}
    \begin{subfigure}[t]{0.41\textwidth}
    \includegraphics[width=\textwidth]{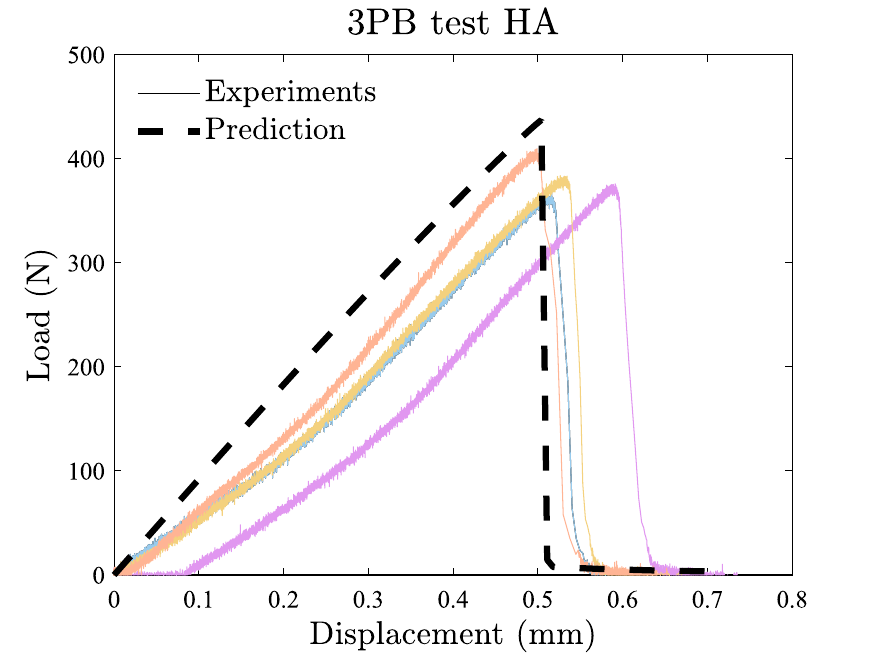}
    \caption{}
    \label{}
    \end{subfigure}\hspace{0.05\textwidth}
    \begin{subfigure}[t]{0.41\textwidth}
    \includegraphics[width=\textwidth]{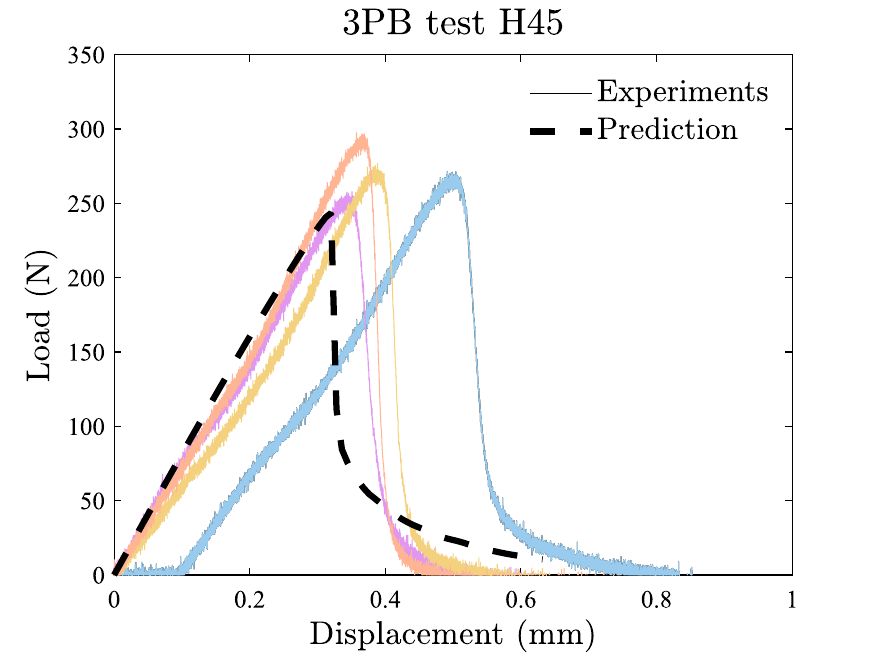}
    \caption{}
    \label{}
    \end{subfigure}
    \caption{Numerical predictions of the force versus displacement response of the three-point bending tests provided as part of the challenge data. Four types of tests have been conducted: (a) HC, (b) HB, (c) HA, and (d) H45. A black dashed line is used for the finite element results and solid colored lines are used for the experimental data consisting of four replicate experiments per testing configuration.}
    \label{fig:Calib_LD}
\end{figure}
\begin{figure}[H]
    \centering
    \includegraphics[width=0.8\textwidth]{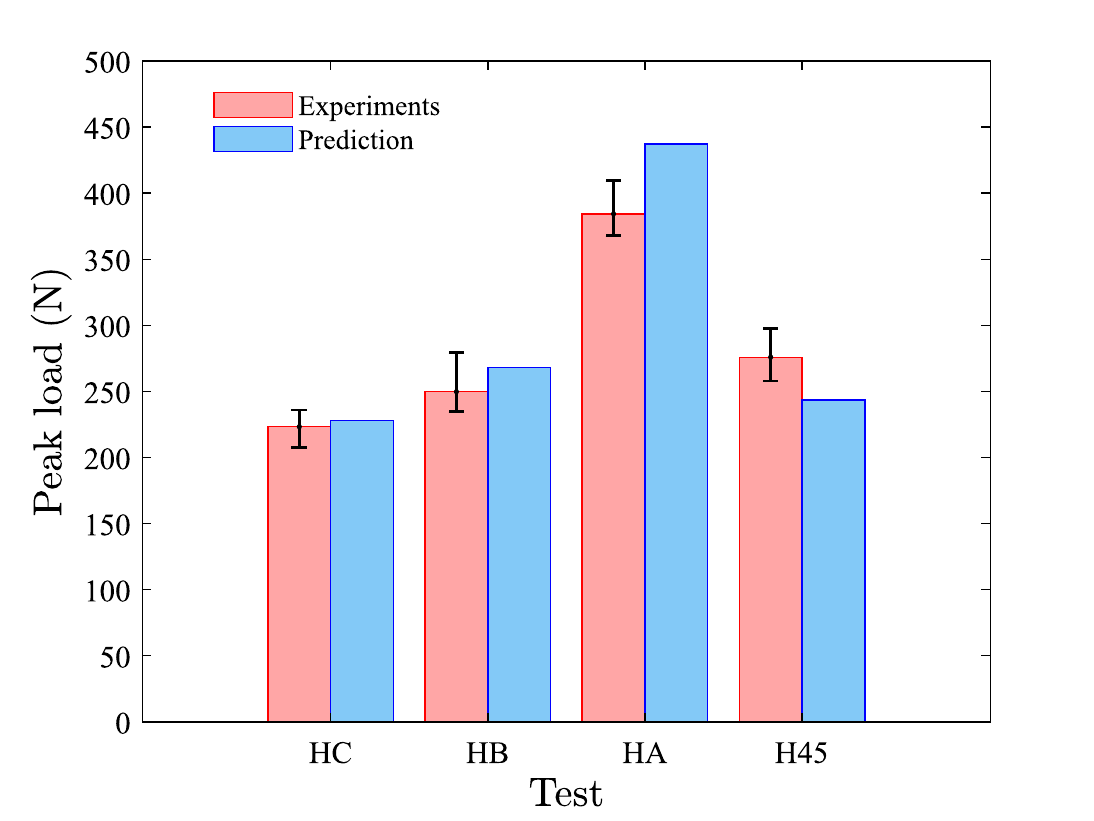}
    \caption{Numerical and experimental estimations of the critical (peak) load for the three-point bending tests provided as part of the challenge data. The experimental data is reported as the average peak load with an error bar.}
    \label{fig:LD_Challenge-Error}
\end{figure}

Numerically predicted force versus displacement responses for the four three-point bending calibration tests are given in Fig. \ref{fig:Calib_LD}, next to the experimental results (4 repeated tests per configuration). The fit is particularly good for the HC case, unsurprisingly as it was used as a calibration benchmark, but a good agreement is overall attained both in terms of peak load and critical displacement. A more quantitative comparison is provided in Fig. \ref{fig:LD_Challenge-Error}, where the peak load is shown for both simulations and experiments. In the latter, the average of the four experiments conducted per test is reported and error bars have been included to quantify the experimental scatter. While the agreement is overall good, numerical simulations of test HA and H45 appear to respectively overestimate and underestimate the peak load. Nevertheless, even in those cases, numerical predictions deviate only $\sim$10\% from the average peak load and the error is less than 7\% from the closest experimental measurement. This level of differences is arguably to be expected considering that the model assumes that the 3D-printed rock is homogeneous and isotropic.

\begin{figure}[H]
    \centering
    \begin{subfigure}[t]{0.308\textwidth}
    \includegraphics[width=\textwidth]{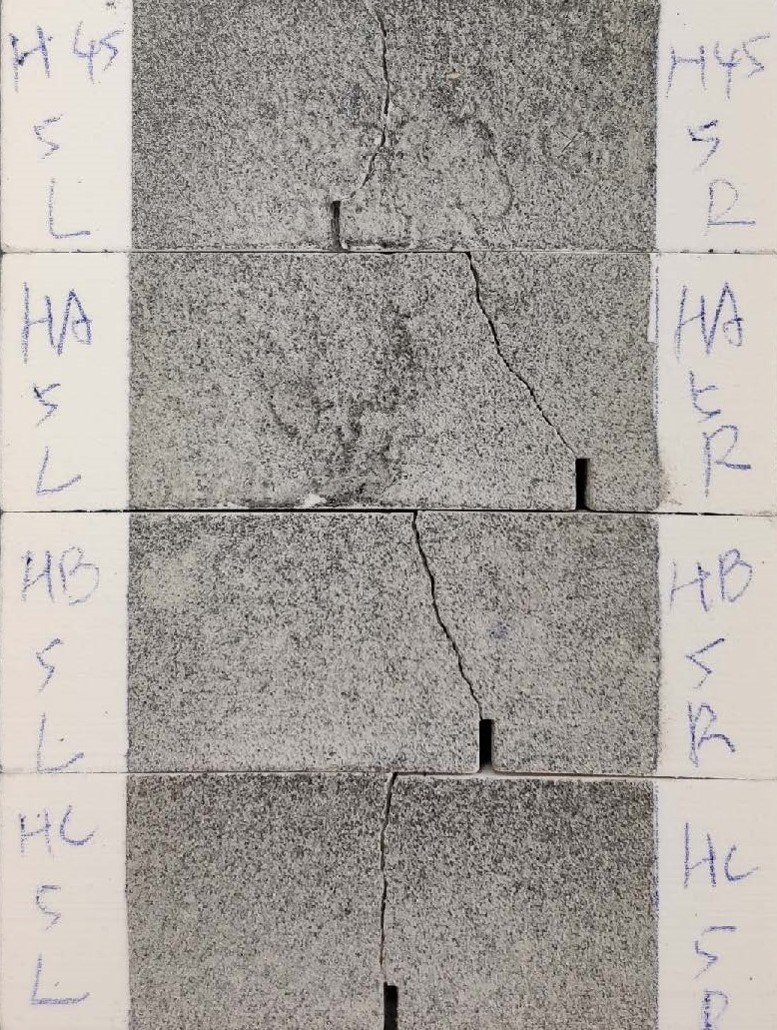}
    \caption{}
    \label{}
    \end{subfigure}
    \begin{subfigure}[t]{0.3\textwidth}
    \includegraphics[width=\textwidth]{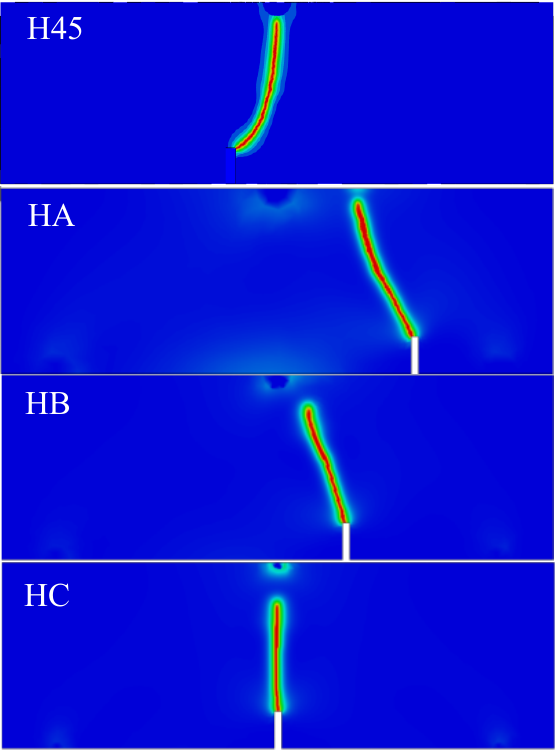}
    \caption{}
    \label{}
    \end{subfigure}
    \begin{subfigure}[t]{0.3\textwidth}
    \includegraphics[width=\textwidth]{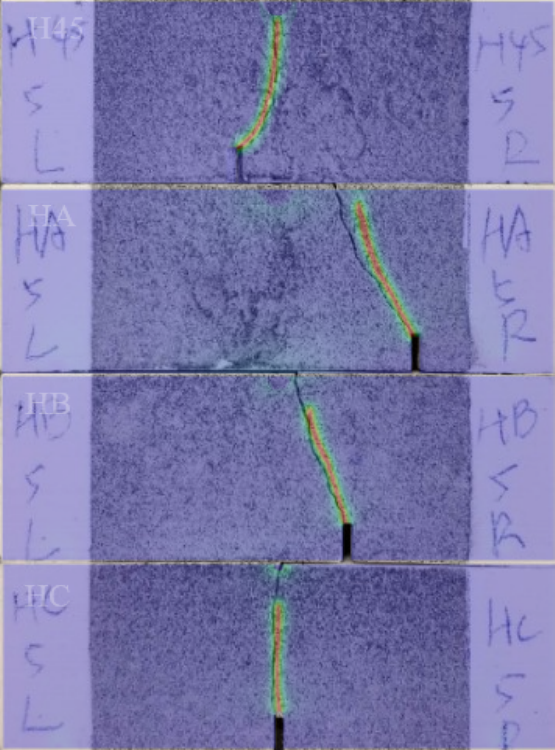}
    \caption{}
    \vspace{5mm}
    \label{fig:Calib_CrackPath-Overlap}
    \end{subfigure}
    \begin{subfigure}[t]{0.8\textwidth}
    \includegraphics[width=\textwidth]{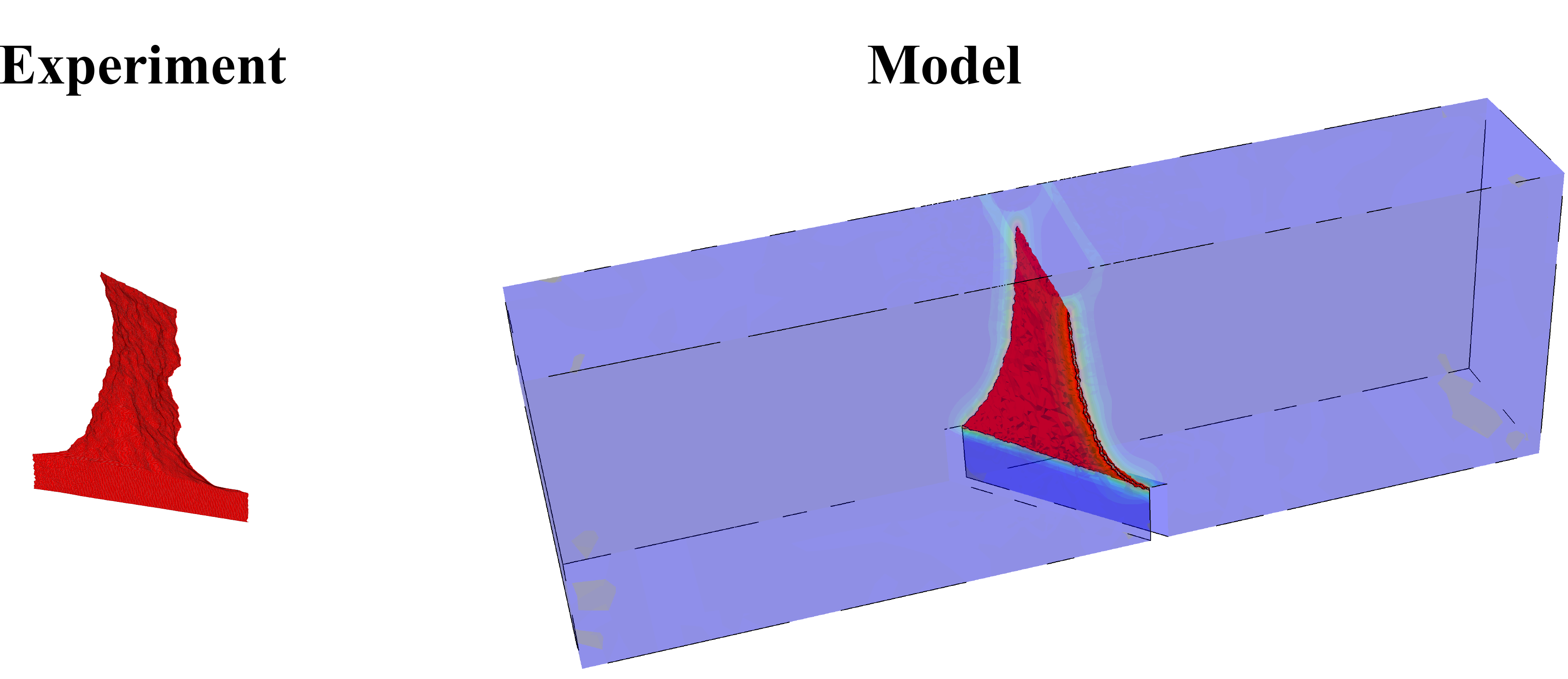}
    \caption{}
    \label{}
    \end{subfigure}
    \begin{subfigure}[t]{0.07\textwidth}
    \includegraphics[width=\textwidth]{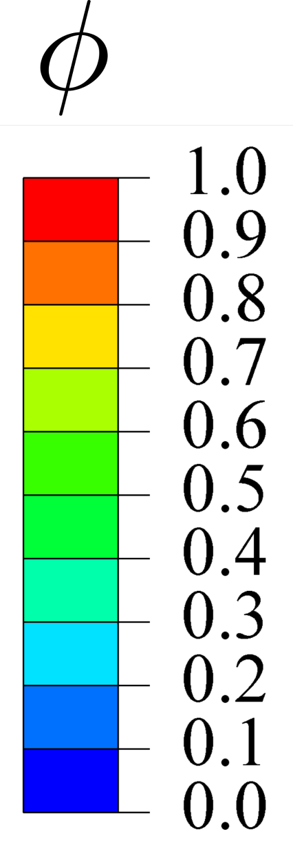}
    \label{}
    \end{subfigure}
    \caption{Numerical and experimental results of crack trajectories and morphology for the three-point bending tests provided as part of the challenge data: (a) experimental crack paths, (b) phase field fracture model predictions, (c) overlap between experimental and numerical results, and (d) comparison of crack morphologies for H45.}
    \label{fig:Calib_CrackPath}
\end{figure}

Next, crack trajectories are compared with DIC measurements. In addition, the predicted crack morphology in the 3D case study (H45) is also compared to the laser profilometry profile. The results are shown in Fig. \ref{fig:Calib_CrackPath}. Fig. \ref{fig:Calib_CrackPath}b shows the predicted 2D crack trajectories, with red colour denoting the regions with $\phi=1$ (i.e., cracks), and Fig. \ref{fig:Calib_CrackPath}c overlaps the experimental (Fig. \ref{fig:Calib_CrackPath}a) and computational (Fig. \ref{fig:Calib_CrackPath}b) results. This overlap reveals an almost perfect agreement between the two, showcasing the ability of the model to capture complex crack trajectories that have not been predefined. A good agreement is also attained in the predictions of crack morphology for the 3D analysis (case H45), as shown in Fig. \ref{fig:Calib_CrackPath}d, with the predicted crack morphology exhibiting the same shape and contortions as the laser profilometry-based measurements. 

\subsection{Blind predictions of the challenge test}
\label{sec:ChallengeRes}


Finally, we use our phase field model to deliver \emph{blind} estimates of force versus displacement behaviour, crack trajectory and crack surface morphology for the challenge test. Our results, submitted to the challenge prior to the release of the corresponding data, are shown here and compared to the outcome of the laboratory tests. It should be noted that calculations were conducted with two types of elements (brick and tetrahedral) and while the results were similar, some differences were found, which are discussed in \ref{App:ElementType}. The results presented in this Section correspond to those calculated using brick elements. After a mesh sensitivity study, the 3D model employed uses a total of 904,429 8-node trilinear brick elements, with the mesh being selectively refined in the regions of potential crack growth (see Fig. \ref{fig:Challenge-Configll}b).\\

\begin{figure}[H]
    \centering
    \begin{subfigure}[t]{.49\textwidth}
    \includegraphics[width=\textwidth]{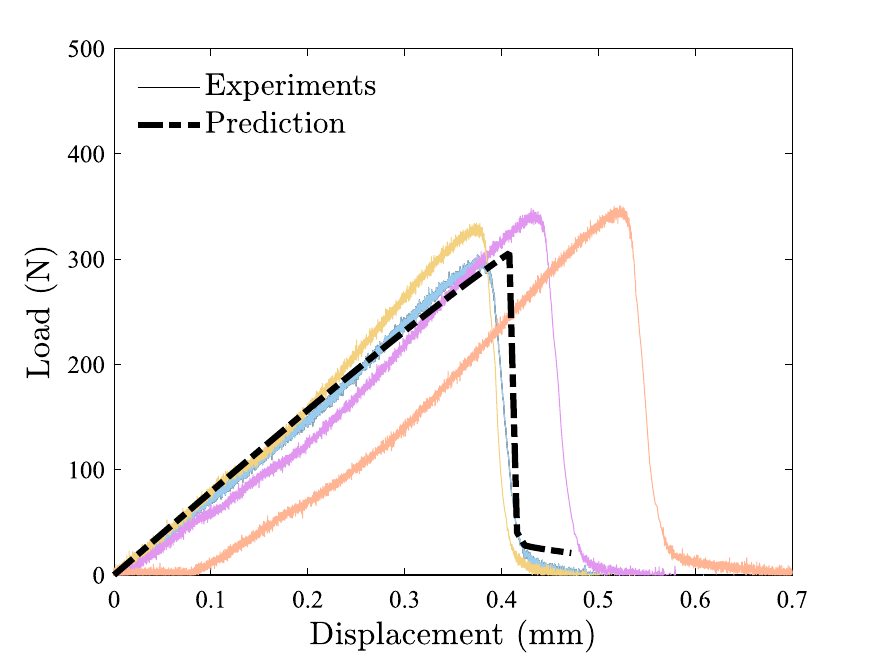}
    \caption{}
    \label{fig:LD_Challenge-a}
    \end{subfigure}
    \begin{subfigure}[t]{.49\textwidth}
    \includegraphics[width=\textwidth]{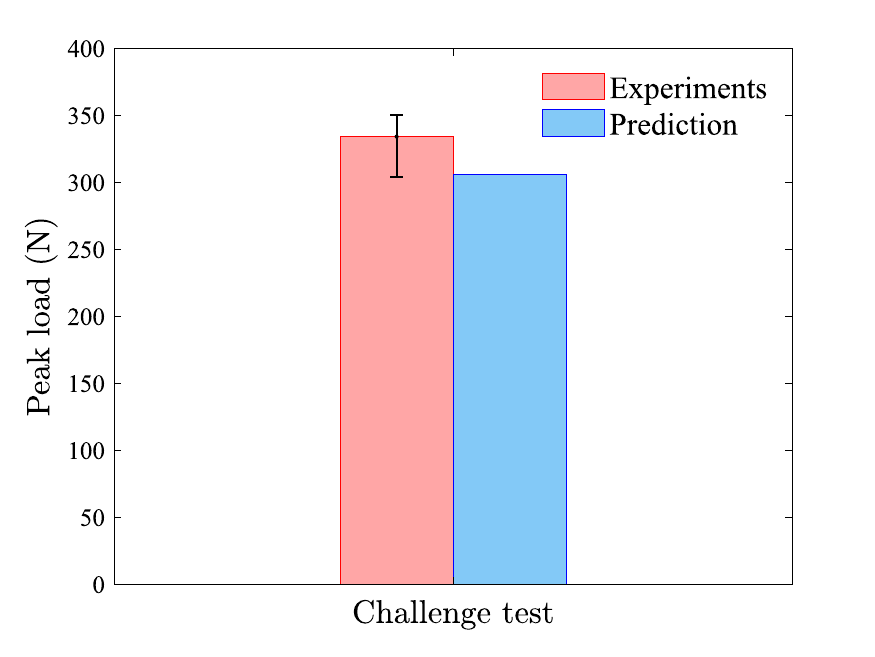}
    \caption{}
    \label{fig:LD_Challenge-b}
    \end{subfigure}
    \caption{Comparison between phase field fracture predictions and experimental data, released a posteriori, for the challenge test: (a) force versus displacement response, showing numerical results (dashed line) and four replicate experiments, and (b) peak load, with the experimental data reported as the average peak load with an error bar.}
    \label{fig:LD_Challenge}
\end{figure}

First, we compare our numerical predictions of the macroscopic force versus displacement response; the results are shown in Fig. \ref{fig:LD_Challenge}. The numerical model appears to succeed in predicting with reasonable accuracy the force versus displacement behaviour, providing a peak load and a critical displacement that lie within the experimental data. As shown in Fig. \ref{fig:LD_Challenge}b, the peak load appears to be slightly lower than the test average but falls within the experimental scatter. 

\begin{figure}[H]
    \centering
    \begin{subfigure}[t]{0.47\textwidth}
    \includegraphics[width=\textwidth]{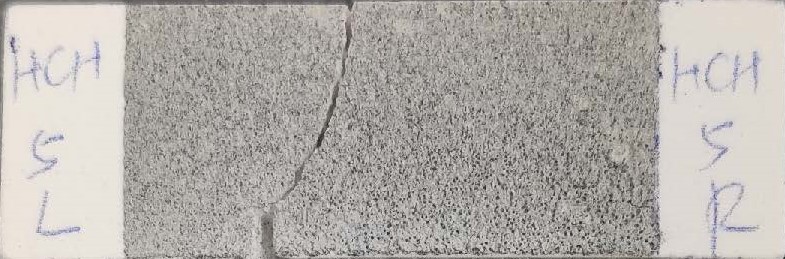}
    \caption{}
    \label{}
    \end{subfigure} \\ 
    \begin{subfigure}[t]{0.47\textwidth}
    \includegraphics[width=\textwidth]{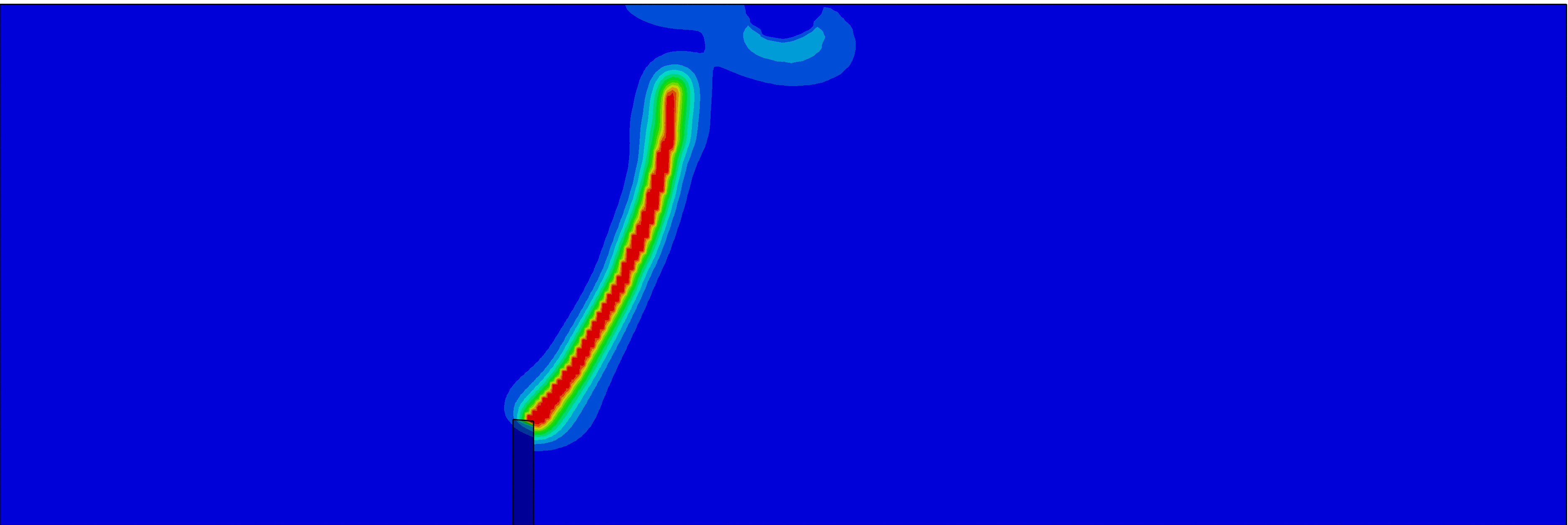}
    \caption{}
    \label{}
    \end{subfigure} \\
    \begin{subfigure}[t]{0.47\textwidth}
    \includegraphics[width=\textwidth]{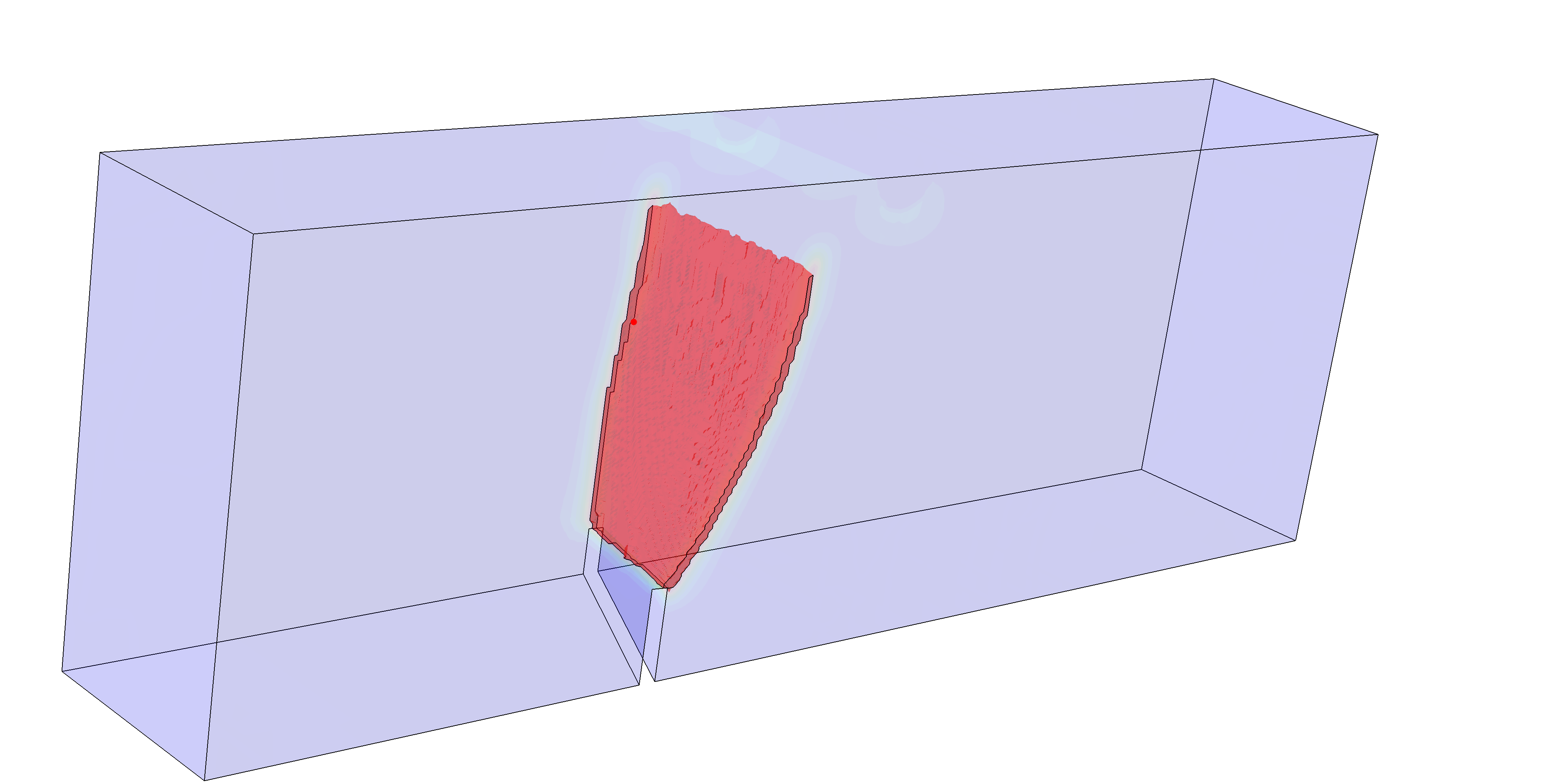}
    \caption{}
    \label{fig:challenge_phi-c}
    \end{subfigure} \\ \hspace{0.05\textwidth}
    \begin{subfigure}[t]{0.4\textwidth}
    \includegraphics[width=\textwidth]{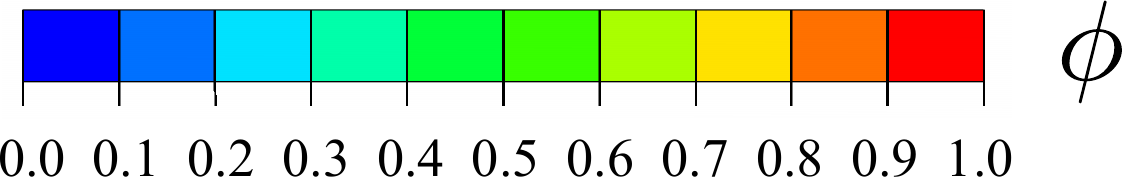}
    \label{}
    \end{subfigure}
    \caption{Crack trajectories. Comparison between phase field fracture predictions and experimental data, released a posteriori, for the challenge test: (a) experimental results, (b) phase field contours, and (c) overlap of numerical and experimental results.}
    \label{fig:challenge_phi}
\end{figure}

The crack path predictions are provided in Fig. \ref{fig:challenge_phi}, as depicted by the phase field $\phi$ contours. The experimental results are also provided (Fig. \ref{fig:challenge_phi}a), together with an overlap of model and experimental crack trajectories (Fig. \ref{fig:challenge_phi}c). The results show a remarkable agreement between computations and experiments, showcasing the ability of the phase field fracture model presented to deliver accurate blind estimations.

\

\begin{figure}[H]
    \centering 
    \begin{subfigure}[t]{0.6\textwidth}
    \includegraphics[width=\textwidth]{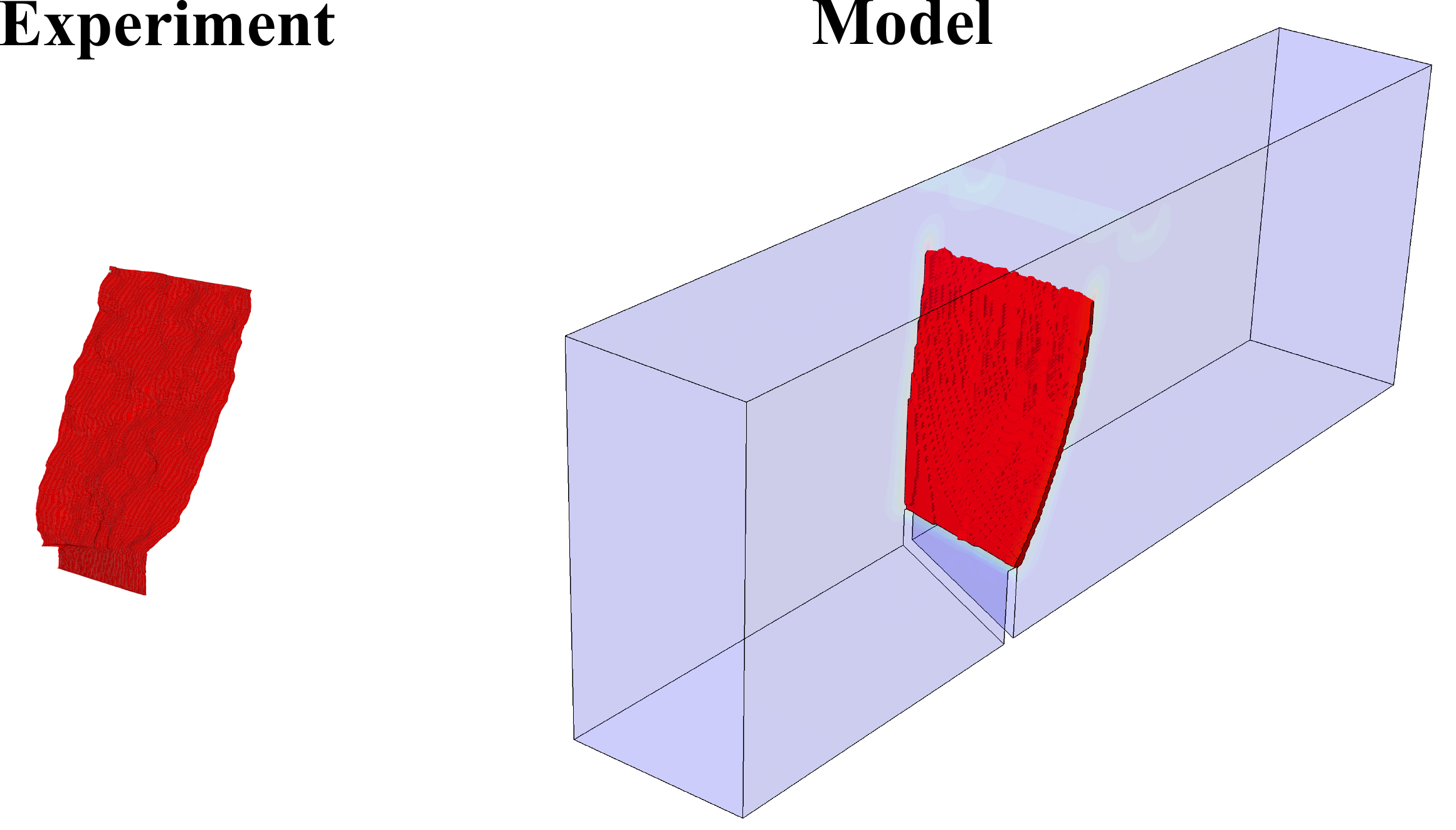}
    \label{Phi-Challenge-3D.png}
    \end{subfigure} 
    \begin{subfigure}[t]{0.07\textwidth}
    \includegraphics[width=\textwidth]{V-Legend.png}
    \label{}
    \end{subfigure}
    \caption{Three-dimensional crack surface morphology. Comparison between phase field fracture predictions and experimental data, as measured with laser profilometry.}
    \label{fig:challenge_frcatureSurface}
\end{figure}

Finally, model predictions are benchmarked against the last piece of data provided: crack surface morphology, as measured using laser profilometry. The experimental data, provided as asperity data for 250 rows and 120 columns in 0.1 mm intervals, is plotted using MATLAB. The results are compared in Fig. \ref{fig:challenge_frcatureSurface}. As can be observed, the crack surface morphology predicted with the phase field model appears to be in good agreement with the experimentally determined crack surface profile, which was released after the submission of the model predictions.

\section{Summary, conclusions and outlook}

We have described our contribution to the \emph{Damage Mechanics Challenge}, which exploited the strengths of the phase field fracture model \cite{Bourdin2008} to deliver \emph{blind} estimates (i.e., predictions submitted before the challenge data was released). Phase field fracture models are grounded on Griffith's energy balance and the thermodynamics of fracture, and accordingly deliver predictions in agreement with the conventional fracture mechanics theory. The aim was to demonstrate the predictive potential of simple, physically-sound models based on a well-established theory. Maximising computational efficiency and simplifying implementation were also targets of this work, which used an unconditionally stable monolithic solution scheme and straightforwardly implemented the model in a commercial finite element package by exploiting the analogy between heat transfer and the phase field evolution equations \cite{Materials2021,AES2021}. 

Our model predictions relied only on four parameters, with a clear physical interpretation: Young's modulus $E$, Poisson's ratio $\nu$, toughness $G_c$ and strength $\sigma_c$, with the last one being defined through the choice of phase field length scale $\ell$ \cite{Tanne2018,PTRSA2021}. Our study showed that a simple, mode I three-point bending test was sufficient to calibrate the model parameters, with the calibrated model accurately predicting the failure characteristics of the other three three-point bending experiments provided as calibration data, which were more intricate and intrinsically mixed-mode. A very good agreement with experiments was observed across all available data: force versus displacement response, crack trajectory and 3D crack morphology. Furthermore, the submitted predictions for the challenge test were shown to deliver a remarkable agreement with the experimental data, released a posteriori. The agreement was found to be excellent across all the data provided: force versus displacement response, crack propagation paths and surface crack profile.\\

The results presented further showcase the ability of phase field fracture models to capture complex cracking phenomena in a physically sound fashion. Only the force versus displacement curve of one conventional, mode I three-point bending test sufficed to calibrate a model that could deliver reliable predictions not only for the challenge test but also for the remaining calibration data, across a wide range of scenarios, data (load carrying capacity, crack paths and morphology) and loading configurations. This is despite the challenge being based on an additively manufactured material (a type of gypsum mortar) that is rare and for which little information is available. Phase field approaches link damage and fracture mechanics, providing the computational robustness of non-local damage models while delivering predictions based on well-established fracture parameters and in agreement with decades of fracture mechanics development and understanding. However, one could envisage ways to complicate the challenge that would have potentially showcased the limitations of phase field fracture modelling and other state-of-the-art computational models. For example, rocks are typically heterogeneous porous materials yet material heterogeneity played a secondary role in this challenge. Within this realm, several classes of heterogeneous rocks have been shown to exhibit distinct mode I and mode II critical energy release rates ($G_{c}^I$, $G_{c}^{II}$). This was not observed in the challenge data, where cracks appear to grow according to the direction of maximum energy release rate. In any case, phase field models have been recently developed to account for material anisotropy \cite{Teichtmeister2017,Bleyer2018} and shear fracture characteristics \cite{Bryant2018,Feng2022a}. Additionally, conventional phase field models assume a failure surface that is symmetric over the tensile and compressive regimes. This was appropriate for this challenge, as the chosen loading configuration (three-point bend testing) resulted in crack growth due to tensile stress states. However, rock-like materials are known to exhibit asymmetric failure surfaces and these would have played a role under more intricate loading conditions. Fortunately, recent years have seen the development of phase field models capable of accommodating arbitrary failure surfaces, such as Drucker-Prager \cite{Lorenzis2021,Navidtehrani2022}. A new edition of this challenge, which could potentially test these complex regimes, would be very welcomed.

\section*{Acknowledgments}
\label{Sec:Acknowledge of funding}

\noindent Y. Navidtehrani acknowledges financial support from the Ministry of Science, Innovation and Universities of Spain through grant PGC2018-099695-B-I00. E. Mart\'{\i}nez-Pa\~neda was supported by an UKRI Future Leaders Fellowship (grant MR/V024124/1). R. Duddu acknowledges the funding support from the National Science Foundation’s Office of Polar Programs via CAREER grant no. PLR-1847173.



\appendix
\setcounter{figure}{0}

\appendix

\section{Finite element implementation of phase field fracture}
\label{App:FEM}

For the sake of facilitating reproducibility, in the following we proceed to describe the characteristics of a finite element implementation of the phase field fracture model. It is emphasised that this information, required for an element-level implementation, is not required if the heat transfer analogy approach is followed, as discussed in Section \ref{Sec:NumImplementation}.\\

We start by taking into account the outlined constitutive selections, including the definition of a history field $\mathcal{H}$, and calculate the stationary of $\mathcal{E}_\ell$ with respect to the primal variables $\mathbf{u}$ and $\phi$; the resultant expression takes the form: 
\begin{align}\label{Eq:weakformApen}
        \partial \mathcal{E}_\ell (\mathbf{u},\phi) = \int_{\Omega}&\biggl\{\left[\left( 1 - \phi \right)^2 \frac{\partial \psi^+_0 \left( \bm{\varepsilon} (\mathbf{u}) \right)}{\partial \bm{\varepsilon} (\mathbf{u})} + \frac{\partial \psi^-_0 \left( \bm{\varepsilon} (\mathbf{u}) \right)}{\partial \bm{\varepsilon} (\mathbf{u})} \right] \delta \boldsymbol{\varepsilon} (\mathbf{u}) -2(1-\phi) \delta \phi \mathcal{H} \nonumber \\
        &+G_{c}\left[\frac{1}{\ell} \phi \delta \phi+\ell \nabla \phi \cdot \nabla \delta \phi\right] - \mathbf{b} \cdot \delta \mathbf{u} \biggr\} \mathrm{d} V -\int_{\partial \Omega} \mathbf{T} \cdot \delta \mathbf{u} \, \mathrm{d} S = 0 \, .
\end{align}

\noindent where $\delta \mathbf{u}$ and $\delta \phi$ are arbitrary fields (test functions). Then, considering the stress definition, Eq. (\ref{eq:Cauchy}), one can reformulate Eq. (\ref{Eq:weakformApen}) into two coupled weak form equations
\begin{align}\label{Eq:weakformApen1}
  \int_\Omega \left\{ \left[\left( 1 - \phi \right)^2 \bm{\sigma_0^+} +\bm{\sigma_0^-} \right] : \delta \bm{\varepsilon} (\mathbf{u}) -\mathbf{b} \cdot \delta \mathbf{u}  \right\} \text{d} V - \int_{\partial \Omega} \mathbf{T} \cdot \delta \mathbf{u}~\mathrm{d} S=0  \, \\
\int_{\Omega} \left\{ {- 2 (1 - \phi)\delta \phi} \, \mathcal{H} +
        G_c \left[ \frac{1}{ \ell}\phi  \delta \phi + \ell \nabla \phi  \nabla \delta \phi \right] \right\}  \, \mathrm{d}V = 0  \, ,
\end{align} 

We define a finite element discretisation to formulate the element stiffness matrix $\bm{K}^e$ and the residual vector $\mathbf{R}^e$. Using Voigt notation, the nodal variables for the displacement field, denoted as $\mathbf{\hat{u}}$, and the phase field $\hat{\phi}$, are interpolated as follows
\begin{equation}\label{eq:Ndiscret}
\mathbf{u} = \sum_{i=1}^m \bm{N}_i \hat{\mathbf{u}}_i, \hspace{1cm} \phi =  \sum_{i=1}^m N_i \hat{\phi}_i \, ,
\end{equation}

\noindent where $N_i$ represents the shape function associated with node $i$, and $\bm{N}_i$ is the shape function matrix. Additionally, $m$ denotes the total number of nodes per element, while $\hat{\mathbf{u}}_i$ and $\hat{\phi}_i$ respectively represent the displacement and phase field at node $i$. In a similar manner, the associated gradient quantities can be discretized using the corresponding \textbf{B}-matrices, which contain the derivatives of the shape functions;
\begin{equation}\label{eq:Bdiscret}
\bm{\varepsilon} = \sum\limits_{i=1}^m \bm{B}^{\bm{\mathrm{u}}}_i \hat{\mathbf{u}}_i, \hspace{0.8cm}  \nabla\phi =  \sum\limits_{i=1}^m \mathbf{B}_i \hat{\phi}_i \, .
\end{equation}

The discretized residuals for each of the primal kinematic variables are then expressed as:
\begin{align}
    & \mathbf{R}_i^\mathbf{u} = \int_\Omega \left\{\left( 1 - \phi \right)^2 \left(\bm{B}^\mathbf{u}_i\right)^T \bm{\sigma}_0^+ + \left(\bm{B}^\mathbf{u}_i\right)^T \bm{\sigma}_0^- \right\} \, \text{d}V \, -\int_{\Omega}\left(\mathbf{N}_{i}^{\mathrm{u}}\right)^{T} \mathbf{b}~ \mathrm{d} V-\int_{\partial \Omega_{h}}\left(\mathbf{N}_{i}^{\mathrm{u}}\right)^{T} \mathbf{T}~ \mathrm{d} S , \\
    & 
\mathbf{R}_{i}^{\phi}=\int_{\Omega}\left\{-2(1-\phi) N_{i} \mathcal{H}+G_{c}\left[\frac{1}{\ell} N_{i} \phi+\ell\left(\mathbf{B}_{i}^{\phi}\right)^{T} \nabla \phi\right]\right\} \mathrm{d} V
\end{align}

The consistent tangent stiffness matrices $\bm{K}$ are then determined by differentiating the residuals with respect to the incremental nodal variables:
\begin{align}
    & \bm{K}_{ij}^{\mathbf{u}} = \frac{\partial \mathbf{R}_{i}^{\mathbf{u}}}{\partial \mathbf{u}_{j}} = \int_{\Omega} \left\{ ( 1 - \phi)^2 (\bm{B}_{i}^{\mathbf{u}})^{T} \bm{C}^+_0 \, \bm{B}_{j}^{\mathbf{u}} + (\bm{B}_{i}^{\mathbf{u}})^{T} \bm{C}^-_0 \, \bm{B}_{j}^{\mathbf{u}}  \right\} \, \text{d}V \, , \\
    & \bm{K}_{ij}^{\phi} = \frac{\partial R_{i}^{\phi} }{ \partial \phi_{j} } =  \int_{\Omega} \left\{ \left( 2 \mathcal{H} + \frac{G_{c}}{ \ell} \right) N_{i} N_{j} +G_{c} \ell   \, \mathbf{B}_i^T\mathbf{B}_j \right\} \, \text{d}V \, ,
\end{align}

\noindent where $\bm{C}^{\pm}_0=\partial \bm{\sigma}^{\pm}_0 /\partial \bm{\varepsilon}(\mathbf{u})$ are the tangent matrices for the positive and negative parts.




\section{Heat transfer analogy}
\label{Sec:HeatApp}

As elaborated in Refs. \cite{Materials2021, AES2021}, one can leverage the analogy with heat transfer to simplify the numerical implementation of the phase field evolution equation in commercial finite element packages. In the presence of a heat source denoted as $r$, the steady-state equation for heat transfer takes the following form:
\begin{equation}\label{Eq:HeatTransfer}
k \nabla^{2} T=-r
\end{equation}

\noindent where, $T$ represents temperature, and $k$ is the thermal conductivity. Equation (\ref{Eq:HeatTransfer}) is analogous to the phase field evolution equation (\ref{eqn:strongForm}b) upon assuming $T \equiv \phi$, $k=1$, and defining the nonlinear heat source $r$ as follows
\begin{equation}
r=\frac{2(1-\phi) \mathcal{H}}{\ell G_{c}}-\frac{\phi}{\ell^{2}}
\end{equation}

Finally, to determine the Jacobian or tangent matrix, we must provide the gradient of the heat source with respect to the phase field (temperature), which reads
\begin{equation}
\frac{\partial r}{\partial \phi}=-\frac{2 \mathcal{H}}{\ell G_{c}}-\frac{1}{\ell^{2}}
\end{equation}

\section{On the influence of the element type}
\label{App:ElementType}

\begin{figure}[H]
    \centering
    \includegraphics[width=0.65\textwidth]{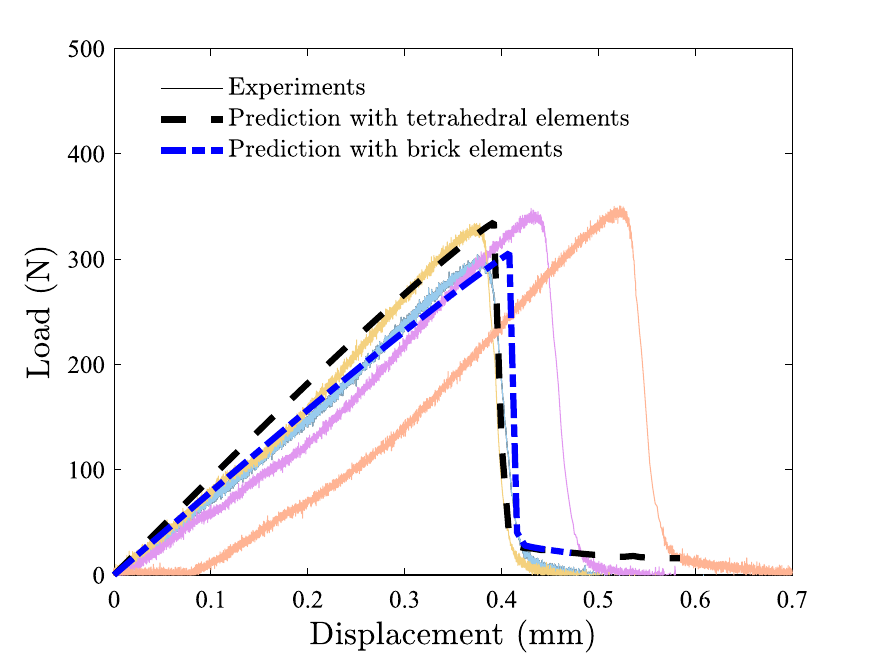}
    \caption{Assessing the role of the element type: load versus force results for the challenge test, as obtained using tetrahedral and brick elements.}
    \label{fig:LD_Challenge1}
\end{figure}

To investigate the sensitivity of model predictions to numerical discretization choices, calculations for the challenge test were conducted using two types of elements, linear tetrahedral elements and linear brick (hexahedral) elements. Overall, a small influence was found, as is to be expected when using sufficiently fine meshes. However, given the size of the 3D models, with millions of degrees-of-freedom (DOFs), an effort was made to use a finite element mesh as coarse as possible outside of the regions of crack growth, and this can result in some sensitivity to the element choice. Thus, to assess this, calculations were conducted with two models, one employing 2,438,970 linear tetrahedral elements (1,673,512 DOFs) and another one employing 904,429 linear brick elements (3,735,636 DOFs). The results obtained are compared in Fig. \ref{fig:LD_Challenge1} in terms of their predicted force versus displacement responses. The experimental results, released a posteriori, are also included. While small, some differences can be observed; the tetrahedral response is stiffer and this leads to a slightly higher peak load. This is to be expected, to a certain extent, as linear tetrahedral elements are known to display stiffer responses if the mesh is not sufficiently fine \cite{puso2006stabilized}. Nonetheless, both sets of numerical results provide a good agreement with experiments. 

\end{document}